\documentclass[longauth]{aa} 
\usepackage{graphicx}
\usepackage{tabularx}
\usepackage{multirow}
\usepackage[colorlinks=true, linkcolor=blue, citecolor=blue]{hyperref}
\usepackage{gensymb}
\usepackage{color}
\usepackage{txfonts}
\usepackage{natbib}
\usepackage{amsmath}
\usepackage{ulem}
\usepackage[flushleft]{threeparttable}
\usepackage[dvipsnames]{xcolor}
\usepackage{rotating}
\usepackage{CJKutf8}
\usepackage{placeins}
\usepackage{booktabs}

\newcommand{\nii}{[N\,\textsc{ii}]}

\newcommand{\oiii}{[O\,\textsc{iii}]}
\newcommand{\ha}{H$\alpha$}
\newcommand{\hb}{H$\beta$}
\newcommand{\sii}{S\,\textsc{ii}}

\begin{document} 

\titlerunning{$\sim$200 Most Massive Galaxies at $z\sim3-15$}
\authorrunning{Xiao et al.}
   \title{A Census of the 200 Most Massive Galaxies Spectroscopically Observed with JWST at $z_{\rm spec} \sim 3-15$}

\author{Mengyuan Xiao\inst{\ref{unige}}\thanks{E-mail: mengyuan.xiao@unige.ch}
\and Pascal A. Oesch\inst{\ref{unige},\ref{dawn},\ref{NBI}}
\and Longji Bing\inst{\ref{Sussex}}
\and Rashmi Gottumukkala\inst{\ref{dawn},\ref{NBI}}
\and Rui Marques-Chaves\inst{\ref{unige}} 
\and Gabriel Brammer\inst{\ref{dawn},\ref{NBI}}
\and Miroslava Dessauges-Zavadsky\inst{\ref{unige}} 
\and David Elbaz\inst{\ref{cea}}
\and Songnan Qi\inst{\ref{unige}} 
\and Anurag Amol Sawarkar \inst{\ref{unige}} 
\and Manuel Aravena\inst{\ref{Santiago},\ref{MINGAL}} 
\and Matthieu Béthermin\inst{\ref{Strasbourg}} 
\and Rachel Bezanson\inst{\ref{Pittsburgh}} 
\and Rychard Bouwens\inst{\ref{leiden}} 
\and Caitlin Casey \inst{\ref{ucsb},\ref{dawn}}
\and Pieter van Dokkum\inst{\ref{yale}} 
\and Andreas L. Faisst\inst{\ref{Caltech}}
\and Yoshinobu Fudamoto\inst{\ref{chiba}}
\and Anna de Graaff\inst{\ref{i_cfa},\ref{i_mpia}}\thanks{Clay Fellow}
\and Olivier Ilbert\inst{\ref{LAM}}
\and Garth Illingworth\inst{\ref{ucsc}} 
\and Guilaine Lagache\inst{\ref{LAM}}
\and Benjamin Magnelli\inst{\ref{cea}}
\and Jorryt Matthee\inst{\ref{Austria}} 
\and Yurina Nakazato\inst{\ref{Flatiron}}
\and Daniel Schaerer\inst{\ref{unige}}
\and Sune Toft\inst{\ref{dawn},\ref{NBI}}
\and Katherine E. Whitaker \inst{\ref{umass},\ref{dawn}} 
\and Christina C. Williams\inst{\ref{NSF}}
}

\institute{Department of Astronomy, University of Geneva, Chemin Pegasi 51, 1290 Versoix, Switzerland\label{unige}
\and Cosmic Dawn Center (DAWN), Denmark \label{dawn}
\and  Niels Bohr Institute, University of Copenhagen, Jagtvej 128, K\o benhavn N, DK-2200, Denmark \label{NBI}
\and Astronomy Centre, University of Sussex, Falmer, Brighton BN1 9QH, UK  \label{Sussex}
\and Universit{\'e} Paris-Saclay, Universit{\'e} Paris Cit{\'e}, CEA, CNRS, AIM, 91191, Gif-sur-Yvette, France \label{cea}
\and Instituto de Estudios Astrof\'{\i}cos, Facultad de Ingenier\'{\i}a y Ciencias, Universidad Diego Portales, Av. Ej\'ercito 441, Santiago, Chile \label{Santiago}
\and Millenium Nucleus for Galaxies (MINGAL) \label{MINGAL}
\and  Université de Strasbourg, CNRS, Observatoire astronomique de Strasbourg, UMR 7550, 67000 Strasbourg, France \label{Strasbourg}
\and Department of Physics and Astronomy and PITT PACC, University of Pittsburgh, Pittsburgh, PA 15260, USA \label{Pittsburgh}
\and Leiden Observatory, Leiden University, PO Box 9500, 2300 RA Leiden, The Netherlands\label{leiden}
\and Department of Physics, University of California, Santa Barbara, CA 93106, USA\label{ucsb}
\and Astronomy Department, Yale University, 52 Hillhouse Ave, New Haven, CT 06511, USA\label{yale}
\and Caltech/IPAC, 1200 E. California Blvd. Pasadena, CA 91125, USA \label{Caltech}
\and Center for Frontier Science, Chiba University, 1-33 Yayoi-cho, Inage-ku, Chiba 263-8522, Japan \label{chiba}
\and Center for Astrophysics $|$ Harvard \& Smithsonian, 60 Garden St., Cambridge MA 02138 USA\label{i_cfa}
\and Max-Planck-Institut f\"ur Astronomie, K\"onigstuhl 17, D-69117 Heidelberg, Germany\label{i_mpia}
\and Aix Marseille Univ, CNRS, CNES, LAM, Marseille, France \label{LAM}
\and Department of Astronomy and Astrophysics, University of California, Santa Cruz, CA 95064, USA\label{ucsc}
\and Institute of Science and Technology Austria (ISTA), Am Campus 1, 3400 Klosterneuburg, Austria \label{Austria}
\and Center for Computational Astrophysics, Flatiron Institute, 162 5th Avenue, New York, NY 10010\label{Flatiron}
\and Department of Astronomy, University of Massachusetts Amherst, Amherst MA 01003, USA\label{umass}
\and NSF National Optical-Infrared Astronomy Research Laboratory, 950 North Cherry Avenue, Tucson, AZ 85719, USA \label{NSF}
}
\date{Received xxx; accepted xxx}

  \abstract
 {Massive galaxies provide strong tests of galaxy formation models, yet a comprehensive spectroscopic view of their properties and demographics in the early Universe has remained elusive. Here we present a JWST spectroscopic census of the 200 most massive galaxies at $z_{\rm spec}\sim3-15$, selected using an evolving stellar-mass threshold motivated by the halo mass function and anchored at $\log(M_{\star}/M_\odot)>10$ at $z\sim5$. These galaxies represent the top 3\% most massive systems among all publicly available NIRSpec/prism observations in the DAWN JWST Archive. 
While not volume-complete, the sample provides the first statistical spectroscopic view of massive galaxies across the first two billion years of cosmic history.
 We derive their physical properties through joint spectral energy distribution (SED) fitting of spectroscopy and photometry, and construct a clean massive galaxy sample after removing little red dots and broad-line AGN contaminants. 
 The inferred stellar masses, and hence the massive-galaxy selection, remain broadly robust under alternative SED-modeling assumptions, including the addition of MIRI photometry. We find that the massive galaxy population evolves strongly with redshift: normal star-forming galaxies (SFGs; dust attenuation of $A_V<1$ mag) dominate at $z\gtrsim6$, while dusty SFGs ($A_V>1$ mag) and quiescent galaxies (QGs) become more common toward lower redshift. 
 Dust attenuation decreases systematically toward higher redshift, consistent with lower levels of dust and metal enrichment in the earliest massive galaxies.
 We identify 29 massive QGs, including a population of recently quenched systems whose star formation declined rapidly within the past $\sim100$ Myr. We further show that both the traditional UVJ and recently proposed $(ugi)_s$ selections suffer substantial inconsistency with the most massive galaxies at $z>3$, motivating a revised $(ugi)_s$ criterion calibrated using our spectroscopic sample.  Finally, we stack the spectra of different massive galaxy populations and cosmic age bins. The inferred formation histories suggest at least two pathways toward quiescence: a dust-enriched pathway linking normal SFGs, dusty SFGs, and QGs, and a more direct pathway connecting normal SFGs and QGs. Massive normal SFGs themselves appear to grow through both relatively gradual and rapid assembly modes. 
 Together, these results suggest that rapid stellar-mass assembly, dust enrichment, and quenching were already shaping the evolutionary pathways of the most massive galaxies within the first billion years after the Big Bang.
 }

   \keywords{galaxies: high-redshift -- 
   galaxies: formation -- 
   galaxies: evolution -- 
   galaxies:  star-formation}
   \maketitle
   
%

\section{Introduction}
Understanding how massive galaxies formed and evolved is a central question in astrophysics. These systems play critical roles in the cosmic baryon cycle: they host the most massive black holes \citep[e.g.,][]{kormendy2013}, enrich the surrounding medium with heavy elements through stellar winds and supernovae \citep[e.g.,][]{maiolino2019}, and regulate star formation via gas accretion and energetic feedback \citep[e.g.,][]{Somerville2015, Naab2017}. As they assemble most of their stars early and rapidly, massive galaxies provide crucial tests for galaxy formation models \citep[e.g.,][]{Neistein2006, Behroozi2019_UNIVERSEMACHINE, Tacchella2018}. In the standard hierarchical galaxy formation models, massive galaxies are expected to grow gradually over time through accretion and mergers within massive dark matter halos \citep[e.g.,][]{White1978, Springel2005Natur, Vogelsberger2020}, often tracing the densest cosmic-web nodes that later evolve into today’s clusters \citep[e.g.,][]{Overzier2016, Chiang2017, Kakimoto2026}.
At the same time, the observed diversity of high-redshift massive systems -- including dusty starbursts, compact quiescent galaxies, and gas-rich mergers 
-- indicates multiple physical pathways shaping their early growth and quenching \citep[e.g.,][]{Elbaz2018, Whitaker2021, Valentino2023}. Determining when and how these galaxies formed, how efficiently they converted gas into stars, and how they eventually quenched, and connect to their dark-matter halos remains a fundamental goal of galaxy evolution studies.

The first four years of the James Webb Space Telescope (JWST) have revolutionized our view of galaxy formation, revealing that galaxies formed \textit{bigger, faster, and earlier} than previously expected \citep[e.g.,][]{Boylan-Kolchin2023, Lovell2023, Dekel2023, Prada2026, Faisst2026, Chworowsky2026}. 
These galaxies, almost as massive as the Milky Way, were found already in place within the first billion years after the Big Bang \cite[e.g.,][]{Xiao2024,Xiao2025_spiral, Lagache2026, Fu2026}. Such highly efficient formation of massive galaxies is indicated among three complementary populations/diagnostics: 1) a large excess at the bright end of the UV luminosity function beyond $z \sim 8-10$ compared to pre-JWST predictions \citep[e.g.,][]{Donnan2023, McLeod2024,McLeod2026, Adams2024, Harikane2024spec, Casey2024, Weibel2025_uvlf}; 2) an overabundance at the massive end of the stellar mass function together with spectroscopically confirmed dusty massive star-forming systems at $z \sim  5-7$ \citep[e.g.,][]{Xiao2024, Xiao2026_S1, Weibel2024, Gentile2024, Chworowsky2024, Shuntov2025, Lagache2026, Bing2024, Bing2025}; 3) the emergence of massive quiescent galaxies at $z>4$ \citep[e.g.,][]{Carnall2023Natur, Valentino2023, Glazebrook2024, deGraaff2025, Weibel2024b, Jizhiyuan2026, ZhangYunchong2026SFH, Whitaker2026_REVIEW, Xiao2026_dormant}, which not only assembled rapidly but also quenched fast.

Beyond their unexpectedly rapid mass assembly and quenching, JWST observations have also revealed that some high-redshift galaxies exhibit surprisingly mature morphologies, including disk-like structures and ordered stellar distributions \citep[e.g.,][]{Kartaltepe2023, Ferreira2023, HuertasCompany2024, HuertasCompany2025, Rowland2024}.
More recently, a morphologically mature massive spiral galaxy at $z\sim5$ has been reported, making it the most distant spiral galaxy candidate to date \citep[Zhúlóng;][]{Xiao2025_spiral}. It exhibits a quiescent bulge and a large, extended star-forming disk, with a morphology similar to that of nearby galaxies, only $\sim$1 Gyr after the Big Bang. Together, these recent findings imply accelerated stellar-mass build-up, early star-formation quenching, and structural transformation within the first billion years. 

In parallel, the release of large public JWST spectroscopic datasets now enables systematic studies of tens of thousands of galaxies with secure redshifts  \citep[e.g.,][]{Oesch2023, Finkelstein2024, Bunker2024, deGraaff2025_rubies, Meyer2025}. 
New JWST/MIRI imaging \citep[e.g., MINERVA;][]{Muzzin2025} extends rest-frame near-IR coverage, improving stellar-mass constraints for the highest-redshift galaxies. 
Altogether, the field has therefore reached a turning point where systematic, spectroscopically anchored studies of early massive galaxies are finally possible; the timing is ideal for a comprehensive study of early massive galaxies.

Leveraging the DAWN JWST archive (DJA)\footnote{\url{https://dawn-cph.github.io/dja}} with currently $\sim$ 10,000 robust JWST NIRSpec/prism spectra at $z>3$ in v4.5, this paper aims to construct a homogeneous spectroscopic catalog of the most massive currently known galaxies from $z\sim3-15$ and to study their physical properties. 
The final catalogs of the massive galaxies (this work) and the parent sample of all prism galaxies \citep{gottumukkala2026}, along with the FIR properties of massive galaxies with ALMA observations (Qi et al. in prep.), will be released on a dedicated webpage.

This paper is organized as follows. In Section~\ref{Sec: data}, we describe the data, joint SED-fitting methodology, and the selection of the massive galaxy sample. Section~\ref{Sect:clean massive} presents the classification of the massive galaxies into different populations. In Section~\ref{Sec: result}, we investigate their demographic, physical, and spectroscopic properties, including their redshift evolution, star-formation activity, photometric selection criteria, and stacked spectral characteristics. We discuss the implications for the formation and evolution of massive galaxies, together with key caveats and systematic uncertainties, in Section~\ref{Sec: evolution}. Finally, our conclusions are summarized in Section~\ref{Sec: conclusion}.

Throughout this paper, we adopt a Chabrier initial mass function \citep[IMF;][]{Chabrier2003} to estimate star formation rate (SFR) and stellar mass ($M_{\star}$). We assume a Planck cosmology \citep{Planck2020} with $(\Omega_\mathrm{m},\, \Omega_\mathrm{\Lambda},\, h)=(0.3,\, 0.7,\, 0.7)$. When necessary, data from the literature have been converted with a conversion factor of $M_{\star}$ \citep[][IMF]{Salpeter1955} = 1.7  $\times$ $M_{\star}$ \citep[][IMF]{Chabrier2003}. All magnitudes are in the AB system \citep{Oke1983}, such that $m_{\rm AB} = 23.9 - 2.5$ $\times$ log(S$_{\nu}$ [$\mu$Jy]).

\section{Data, method and sample}\label{Sec: data}

\subsection{NIRSpec prism spectroscopy and photometry}\label{Sect:spectra data} 
To build a JWST spectroscopic sample of massive galaxies, we select all sources with secure spectroscopic redshifts $z_{\rm spec} > 3$ that have both NIRSpec prism spectroscopy and NIRCam imaging.
We begin with all publicly available prism spectra from the DJA, using the latest v4 reductions \citep{msaexp, deGraaff2025_rubies,Heintz2024,Heintz2025, Pollock2026}\footnote{Latest DJA public spectra: \url{https://s3.amazonaws.com/msaexp-nirspec/extractions/nirspec_public_v4.5.html}}. Only spectra with robust redshift measurements (quality grade = 3) are retained. 

All selected sources have matched photometry from the multi-wavelength catalog, constructed following \citet{Valentino2023}\footnote{Latest photometric catalog database: \url{https://s3.amazonaws.com/grizli-v2/JwstMosaics/v7/index.html?search=phot.fits}} using the latest datasets available in DJA, including HST and JWST/NIRCam photometry for all fields.
For JWST/MIRI, we use available photometry from COSMOS-Web, SMILES, MIDIS, and MINERVA (see Appendix~\ref{appendix1} for more details). The combined photometry covers the rest-frame UV-to-NIR, enabling uniform SED modeling for the entire sample.

\begin{figure*}
\centering
\includegraphics[width=18.5cm]{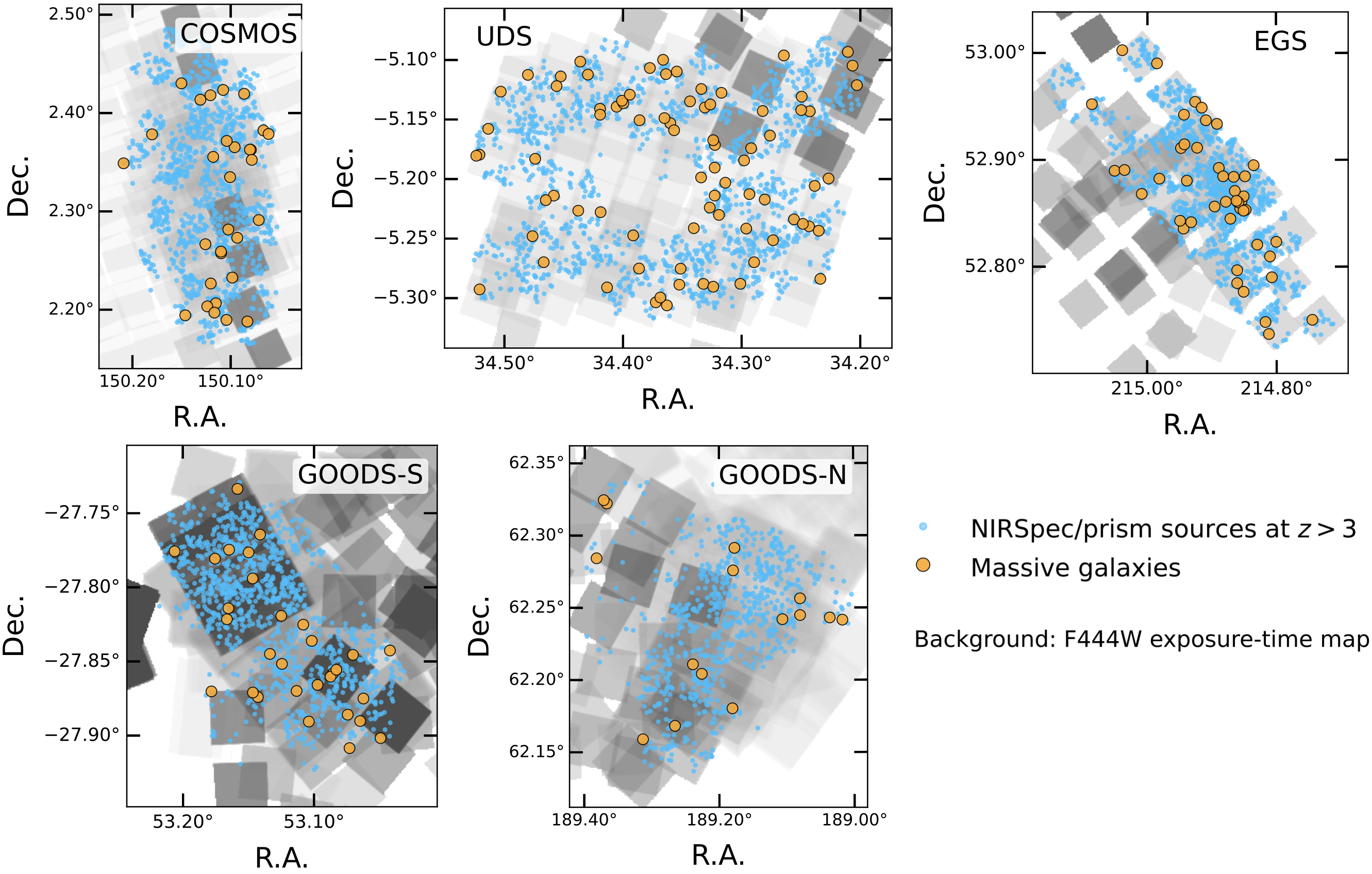}
\caption{\textbf{Sky distribution of the five JWST blank-field legacy surveys: COSMOS, UDS, EGS, GOODS-South, and GOODS-North.} The grey-scale background shows the JWST/F444W exposure-time maps. Blue and orange symbols denote the location of all publicly available NIRSpec/prism sources at $z>3$ and the sample of the most massive galaxies studied in this work, respectively.}
         \label{fig1}
\end{figure*}

The final spectroscopic-photometric catalog spans all prime JWST\ blank legacy fields, including COSMOS, UDS, EGS, GOODS-South, and GOODS-North (Fig.~\ref{fig1}). It contains 6,312 unique galaxies at $z_{\rm spec} > 3$ (7,045 individual prism observations) as of the latest release of DJA v4.5 on February 23, 2026.

\subsection{Joint SED fitting of photometry and prism spectra}\label{Sect:SED} 

To identify massive galaxy candidates and best constrain their star formation histories (SFHs) and stellar populations, we perform joint UV-to-NIR SED fitting by combining the JWST/NIRSpec prism spectra (wavelength coverage of 0.5 - 5.5$\mu$m) and NIRCam broad-band photometry (F090W, F115W, F150W, F200W, F277W, F356W, F444W). This joint analysis leverages the complementary strengths of the two datasets: the broad-band photometry provides broad wavelength coverage and high signal-to-noise ratio (S/N); the prism spectra provide continuous low-resolution spectroscopy, capturing the continuum shape and strong emission lines when present.

We use the \texttt{Bagpipes} code \citep{Carnall2018, Carnall2019} to fit the observed spectra and photometry simultaneously at fixed $z_{\rm spec}$. 
We adopt a non-parametric SFH using a flexible continuity prior following \cite{Leja2019}. The SFH is modeled using ten time bins. The first five bins are fixed at [0, 3, 10, 25, 50, 100] Myr, while the remaining five are logarithmically spaced between 100 Myr and the maximum allowed age, defined as the age of the Universe relative to its age at $z=30$. 
We use the BPASS stellar population models, and apply the \citet{Salim2018} dust attenuation curve, which allows for variable attenuation slopes and UV bump strength. 
The dust attenuation to the rest-frame V-band ($A_{\rm V}$) values is allowed to vary from 0 to 6 mag, the deviation from \cite{Calzetti2000} slope varies from $-1.0$ to 0.3, and the 2175 \AA\, bump strength varies from 0 to 5.  The metallicities are sampled log-uniformly from 0.01 to 2.5 $Z_{\odot}$ and the ionization parameters log$U$ between -4 and -1.  
To correct residual slit-loss of the spectra compared to the photometry, we adopt a second-order polynomial spectral calibration model with Bayesian priors, and introduce a wavelength-independent noise scaling factor with a wide log-uniform prior to absorb any residual systematics. All fits are performed using the \texttt{nautilus} nested sampling algorithm \citep{Lange2023}, which efficiently explores the high-dimensional posterior space.
We have applied this modeling methodology, which we note is similar to and extends upon that presented by \cite{Covelo-Paz2025}, to the full DJA compilation of prism spectra. Additional technical details, as well as a full release of the model products and catalogs (12,275 prism spectra at $z>2$), are provided by \cite{gottumukkala2026}\footnote{\url{https://doi.org/10.5281/zenodo.22734795}}.

 For each galaxy, we obtain posterior distributions for physical parameters including $M_{\star}$, SFR, mass-weighted age, $A_{\rm V}$, and SFH shape. The resulting $M_{\star}$ is used to define our massive galaxy sample (Sect.~\ref{Sect:selection}). Examples of the joint SED fitting results are shown in Appendix~\ref{appendix_sed}.

\subsection{Massive galaxy sample}\label{Sect:massive sample}
\subsubsection{Sample selection}\label{Sect:selection}
Our goal is to trace the build-up and evolution of the most massive galaxies across cosmic time. To construct a representative sample of such systems over a wide redshift range, we do not adopt a fixed stellar-mass threshold, since stellar mass evolves significantly with redshift. Furthermore, although the stellar mass function provides a statistical view of the distribution of galaxy stellar masses \citep[e.g.,][]{Weibel2024, Shuntov2025, Harvey2025}, it is primarily derived from large photometric samples and remains highly uncertain at the massive end, particularly at high redshift. 
Contaminants such as little red dots can lead to overestimated stellar masses \citep[e.g.,][]{Labbe2023a}, while sources with broad photometric-redshift probability distributions, such as the so-called Schrödinger’s galaxies \citep[e.g.,][]{Naidu2022b, Zavala2023}, can scatter across both redshift and stellar-mass bins, further biasing the high-mass end. 
Consequently, the observed stellar mass function does not provide a robust basis for defining a redshift-dependent cutoff.

\begin{figure*}[!htbp]
\centering
\includegraphics[width=15cm]{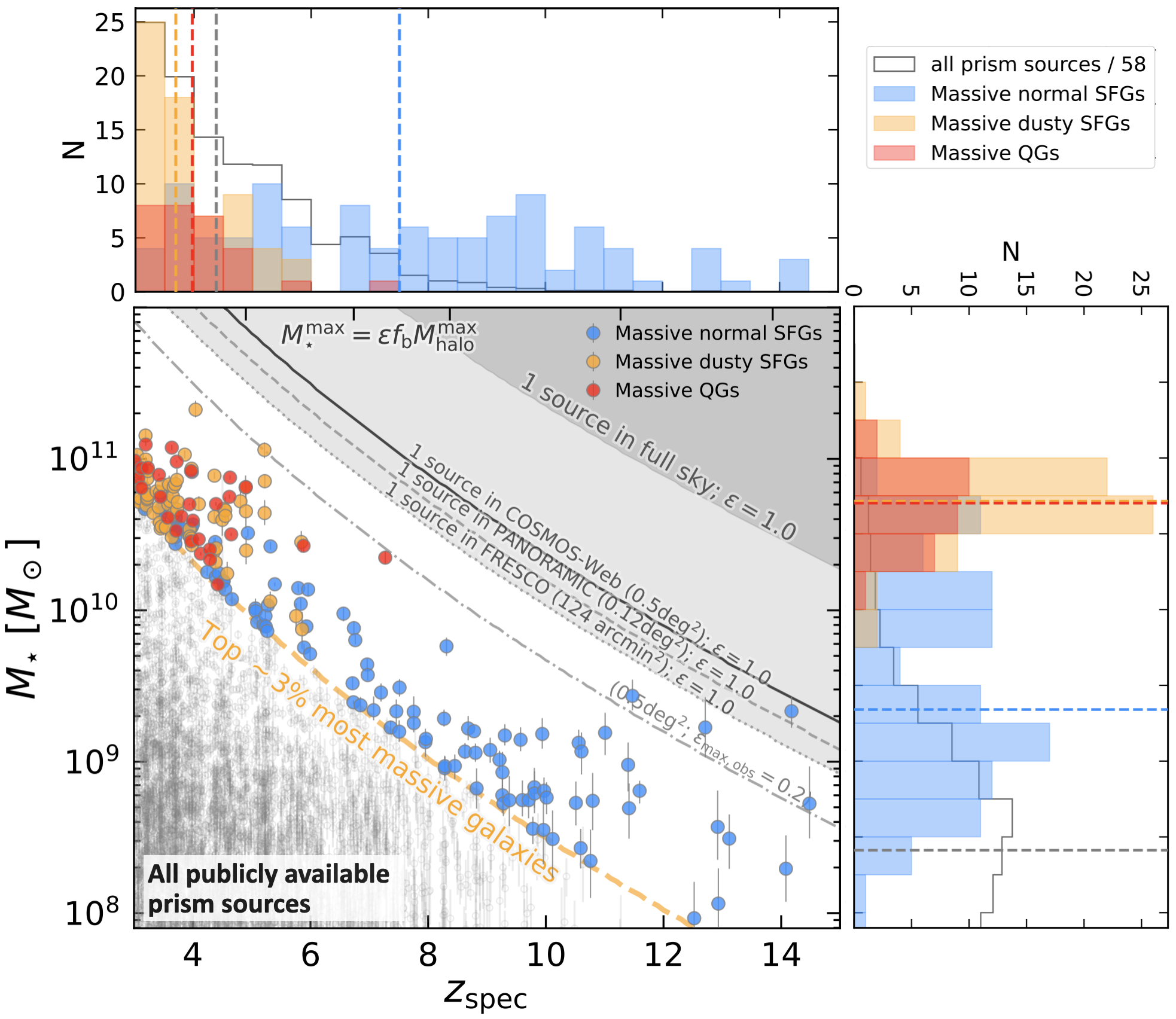}
\caption{\textbf{Stellar mass vs. $z_{\rm spec}$}. 
The 200 most massive galaxies are classified as QGs (red), dusty SFGs (orange), and normal SFGs (blue), following the criteria described in Sect.~\ref{Sect:clean massive}. They represent the top $\sim$3\% most massive galaxies among all 6,312 galaxies (grey open points) at $z_{\rm spec}>3$ with JWST/NIRSpec prism observations in the legacy fields. 
The orange dashed curve shows the evolving stellar-mass threshold used to select massive galaxy candidates (Sect.~\ref{Sect:selection}), motivated by the redshift evolution of the halo mass function and normalized to $\log(M_\star/M_\odot)=10$ at $z=5$.
The dotted, dashed, and solid curves, together with the grey shaded region, indicate the maximum stellar masses corresponding to 100\% baryon-to-star conversion efficiency in the FRESCO, PANORAMIC, COSMOS-Web, and full-sky volumes per $\Delta z=1$, respectively. The dash-dotted line shows the corresponding limit for a 20\% baryon-to-star conversion efficiency in the COSMOS-Web field. The top and right panels show the redshift and stellar-mass distributions of the different populations, respectively, with dashed lines indicating their median values.
We caution that, at $z\gtrsim10$, stellar-mass estimates become increasingly uncertain, as most sources lack photometric coverage redward of F444W\protect\footnotemark (tests see Appendix~\ref{appendix1}, \ref{appendix2}).}
         \label{fig2}
\end{figure*}

Instead, we adopt a simple and physically motivated approach based on the evolution of the dark matter halo mass function \citep{Tinker2008}, inspired by abundance matching \citep[e.g.,][]{Moster2013,  Behroozi2019_UNIVERSEMACHINE}. Under the assumption that the most massive galaxies at each epoch reside in the most massive dark matter halos, we use the cumulative halo mass function to trace the redshift evolution of halo masses at a fixed comoving number density. Assuming a constant baryon-to-star conversion efficiency, this evolution defines the slope of our evolving stellar-mass cut. The overall normalization is then fixed by a representative anchor point, log$(M_{\star}/M_\odot)>10$ at $z\sim5$ (similar to log$(M_{\star}/M_\odot)>9$ at $z\sim8$),  which effectively sets the intercept of the curve. This yields a well-defined redshift-dependent mass cut (see Fig.~\ref{fig2}). With this criterion, we identify a total of 283 candidate massive galaxies from our sample of 6,312 spectroscopically confirmed sources at $z > 3$.

Among 283 candidate massive galaxies, some of these stellar mass estimates may be biased due to contamination from broad-line active galactic nuclei (AGNs), including quasars and the so-called Little Red Dots \citep[LRDs; e.g.,][]{Matthee2024}. These sources can significantly boost the photometric fluxes and alter the continuum shape, leading to unreliable SED fitting results, especially since we do not include any AGN template in the fitting \citep[e.g.,][]{Williams2024_RED, Wangbingjie2024}. Narrow-line AGNs, on the other hand, generally have minimal impact on the broad-band SED and are not expected to bias the derived stellar masses significantly. Therefore, we perform a detailed cleaning procedure to identify and remove the broad-line AGN contaminants, as described in the following sections (Sect.~\ref{Sect:fit} and Sect.~\ref{Sect:agn}).

\footnotetext{The source JADES-GS-z14-0 at $z_{\rm prism}=14.1754$ has $\log M_\star/M_\odot = 9.3\pm0.1$, slightly higher than reported by \cite{Helton2025}. This difference is likely due to contamination from a nearby source, given the $0\farcs25$ radius aperture adopted in the photometric catalog \citep{Valentino2023}. Since the stellar-mass difference remains within the 1$\sigma$ uncertainty, we retain our derived value for consistency with the rest of the sample.} 

\subsubsection{Measurements of spectral properties} \label{Sect:fit}

We performed emission-line fitting for a sample of 283 candidate massive galaxies observed with JWST/NIRSpec prism spectroscopy. For each galaxy, we fit up to five prominent rest-frame optical emission lines: \ha\, (blended with \nii), [S\,\textsc{ii}]~$\lambda\lambda6718,6732$ (modeled as a single component centered at their mean wavelength), H$\beta$, [O\,\textsc{iii}]~$\lambda4960$, and [O\,\textsc{iii}]~$\lambda5008$, using single-component Gaussian profiles. The [O\,\textsc{iii}]~$\lambda4960$/$\lambda5008$ flux ratio was fixed at 1:2.96. The $z_{\mathrm{spec}}$ was allowed to vary within $\pm1\%$ of the DJA catalog value during the fitting process. 

Emission lines were grouped into two configurations: (1) \ha+\sii\, and (2) \hb+\oiii. The local continua were modeled separately using linear fits to line-free regions extending $\pm400$\,\AA\ around each emission-line configuration and subtracted before fitting. Then, the emission-line fitting was carried out using a weighted least-squares method via \texttt{scipy.optimize.curve\_fit}, with inverse-variance weighting based on the flux uncertainties. Initial guesses for line amplitudes were set based on the observed flux range after continuum subtraction, and the Gaussian widths were initialized based on the wavelength-dependent instrumental resolution\footnote{Dispersion curves for the NIRSpec dispersers: \url{https://jwst-docs.stsci.edu/jwst-near-infrared-spectrograph/nirspec-instrumentation/nirspec-dispersers-and-filters\#gsc.tab=0}}. Parameter bounds allowed line amplitudes to vary from 0 to twice the approximate peak value, and line widths from half to twice the instrumental resolution. From the best-fit models, we derived integrated line fluxes, rest-frame equivalent widths (EW$_{0}$), and total line widths (full widths at half maximum, i.e., FWHMs;  including instrumental broadening). Flux uncertainties were estimated from the covariance matrix of the fits, while FWHM uncertainties were derived using Monte Carlo resampling. Emission lines with S/N below 5 were excluded from the final measurements to ensure robustness.

In addition to emission lines, we also modeled the continuum shape by fitting a power-law ($f_\lambda = a \cdot \lambda_{\rm rest}^\beta$) to two broad rest-frame windows: 1500-3200\AA\ and 4200-7000\AA. These regions lie blueward and redward of the Balmer break at 3645\AA, and are used to determine the rest-frame UV ($\beta_{\mathrm{UV}}$) and optical continuum slopes ($\beta_{\mathrm{opt}}$), respectively. To minimize contamination from emission lines, we masked $\pm$50\AA\ around all strong transitions. Fits were performed only if at least 25 valid spectral points remained in a given window after masking. The $\beta_{\mathrm{UV}}$ and $\beta_{\mathrm{opt}}$ provide a quantitative view of the continuum shape and are used in the following Sect.~\ref{Sect:agn} to identify LRDs with “V-shaped” spectra, a feature often linked to broad-line AGNs \citep[e.g.,][]{Labbe2023b, Greene2024, Kocevski2024,Xiao2025_lrd, Hviding2025}.

\subsubsection{Identification of broad-line AGNs and LRDs} \label{Sect:agn}
To remove unreliable stellar mass estimates caused by AGN contamination, we identify broad-line AGNs using three complementary approaches: 
(1) broad Balmer line widths from spectral fitting, (2) compactness combined with a V-shaped continuum, and (3) cross-matching with previously identified AGN/LRD candidates from the literature and performing a visual inspection. Note that our goal is not to compile a complete massive galaxy catalog -- particularly since our sources originate from various JWST programs with different selection biases -- but to build up a clean sample free from sources that could bias the inferred stellar masses. Therefore, any source that satisfies at least one of these criteria is considered an AGN contaminant and removed from our final sample. 

First, we flag galaxies exhibiting broad Balmer emission lines, defined as having a ratio of total to instrumental FWHM greater than 1.5 for either H$\alpha$ or H$\beta$ (depending on prism coverage at a given redshift). At $z \gtrsim 7.2$, however, H$\alpha$ lies beyond the prism wavelength range, and at $z \gtrsim 10$, both H$\alpha$ and H$\beta$ fall outside the spectral coverage. More importantly, many broad-line AGNs may remain unidentified due to $i)$ the limited prism resolution -- particularly at lower redshifts, where the instrumental FWHM reaches $\sim3800$ km s$^{-1}$ for H$\alpha$ at $z\sim3$ and $\sim1800$ km s$^{-1}$ for H$\alpha$ at $z\sim5$; or $ii)$ their Balmer lines which are intrinsically weak and thus not significantly detected. Thus, relying solely on this criterion is insufficient. 

Second, we identify additional candidates using photometric and spectroscopic criteria indicative of LRDs \citep[e.g.,][]{Xiao2025_lrd, Kocevski2024, Hviding2025}. Specifically, we select compact sources with "v-shaped" continuum slopes, defined by:
\begin{eqnarray*}
    && \text{S/N(F444W)} > 14\quad \&\quad \text{m}_{\text{F444W}} < 27.7 \text{ mag},\\
    && \beta_{\rm UV} < -0.2 , \\
    && \beta_{\rm opt} > 0 , \\
    && \text{compact}=f_{\rm F444W}(0\farcs4)/f_{\rm F444W}(0\farcs2) < 1.7.
\end{eqnarray*}
These criteria are motivated by recent studies showing that such sources often host broad-line AGNs \cite[e.g.,][]{Greene2024}. 

Third, we cross-match our candidate massive sources with representative samples of broad-line AGN and LRD candidates identified in the literature studies \citep[][]{Kocevski2024, Hviding2025, Kokorev2024,deGraaff2025_sample, Juodzbalis2026}, which have either medium-resolution NIRSpec spectroscopy or use photometric selection as a complement to our criteria. This step mitigates the limitations of the prism, including its low spectral resolution and redshift-dependent wavelength coverage in both emission lines and continuum. Finally, we also visually inspect all candidate sources to ensure that no obvious AGN contaminants -- such as point-like morphologies, broad features, or quasar-like continua -- were missed. We did not identify any additional LRD or broad-line AGN contaminants missed by the three criteria above.

In total, we identify 83 sources as broad-line AGN contaminants. After removing these, our final clean sample consists of 200 massive galaxies at $z_{\rm spec} \sim 3-15$.

\subsubsection{A representative sample of top 3\% most massive galaxies at $z\sim3-15$ }\label{Sect:test}

Fig.~\ref{fig2} presents the redshift-stellar mass distribution of our final sample of 200 massive galaxies, selected from the full JWST/NIRSpec prism spectroscopic catalog of 6,312 sources at $z_{\rm spec} > 3$ in the legacy field. These galaxies represent approximately the top 3\% most massive systems among all the public prism sources in the first 2 billion years of the Universe.

To place these galaxies in context, we compare their stellar masses to the maximum expected values within various cosmic volumes, including the full sky, COSMOS-Web \citep[0.5 deg$^2$;][]{Casey2023}, PANORAMIC \citep[0.12 deg$^2$;][]{Williams2024}, and FRESCO \citep[124 arcmin$^2$;][]{Oesch2023}, assuming a redshift slice of $\Delta z=1$. These limits are derived from the halo mass function \citep{Tinker2008} and the cosmic baryon fraction ($f_b = 0.158$; \citealt{Planck2020}), adopting a baryon-to-star conversion efficiency $\epsilon$ such that $M_{\star}^\mathrm{max} = \epsilon f_{\rm b} M_\mathrm{halo}^\mathrm{max}$. We illustrate two scenarios: a maximum observational efficiency at lower redshift ($\epsilon_{\rm max,obs} = 0.2$; also the value of the Milky Way; e.g., \citealt{Moster2013, Tacchella2018}) and the theoretical upper limit ($\epsilon = 1$).

Since our sample is taken from a heterogeneous collection of public JWST/NIRSpec prism programs, the effective survey volume cannot be precisely determined. For comparison, we examine the representative surveys listed above and find that some extremely massive galaxies at $z>9$ could lie at $\epsilon=1$, if they were located in smaller-area fields such as FRESCO and PANORAMIC. However, we note that: 1) the inferred efficiency is highly sensitive to the assumed survey volume \citep[and also subject to cosmic variance; see][]{Xiao2024}; 2) at $z\gtrsim9$, stellar-mass estimates become increasingly uncertain, as most sources lack photometric coverage redward of F444W (rest-frame B-band). 
Crucially, none of the 6,312 prism spectroscopic sources at $z_{\rm spec}>3$ exceed the theoretical maximum efficiency of $\epsilon = 1$ across the full sky. 
This indicates that all galaxies observed with JWST/NIRSpec prism to date — including the most massive ones — remain consistent with theoretical expectations under the $\Lambda$CDM framework.

\subsection{Robustness of stellar mass estimates}\label{Sect:robust}
Since stellar mass is the fundamental quantity used both to construct our massive galaxy sample and to derive the physical properties presented throughout this work, we assess the robustness of the inferred stellar masses against a range of commonly adopted SED modeling assumptions. Using the full spectroscopic sample of 6,312 galaxies and the 200 most massive galaxies, we compare stellar masses derived (1) with and without MIRI photometry, (2) using photometry only versus joint photometry+prism fitting, (3) BPASS versus BC03 stellar population synthesis models, (4) the \citet{Salim2018} versus \citet{Calzetti2000} dust attenuation curves, and (5) three alternative star formation histories (delayed-$\tau$, double-power-law+burst, and constant SFH). The detailed comparisons are presented in Appendix~\ref{appendix1} and Appendix~\ref{appendix2}.

Overall, the inferred stellar masses are broadly consistent across the different fitting configurations (Tables~\ref{tab:robustness} and ~\ref{tab:robustness_massive}). The following values refer to the full spectroscopic sample and the 200 massive galaxies, respectively.
The largest deviation is found for the constant-SFH model, which yields both the largest systematic offset ($-0.120$ and $-0.351$ dex) and the largest scatter (NMAD = 0.251 and 0.225 dex). This is expected because a constant SFH cannot reproduce the bursty growth and rapid quenching observed in high-redshift galaxies and is therefore included only as a simplified limiting case rather than as a physically realistic model. Excluding this case, the median offsets are small (0.004-0.046 dex and $-0.152$ to $0.075$ dex), while the NMAD ranges from 0.033 to 0.186 dex and from 0.110 to 0.182 dex.

The inclusion of MIRI photometry has a negligible effect on the inferred stellar masses in general, with a median offset of only 0.004 dex (0.075 dex) and an NMAD of 0.033 dex (0.182 dex) for all prism sources with MIRI (massive sources in this work). However, the relatively large scatter (NMAD = 0.182 dex) of the massive galaxy sample indicates that, although the overall statistical trends remain robust, individual stellar-mass estimates should still be treated with caution. 
More significant discrepancies are primarily associated with likely contaminants, including LRDs and broad-line AGNs (which are excluded from our final massive-galaxy catalog). In comparison, fitting only the photometric data yields a similarly small median offset (0.030 dex) but a substantially larger scatter (NMAD = 0.186 dex), highlighting the importance of prism spectroscopy for reducing degeneracies among stellar mass, stellar age, dust attenuation, and nebular emission.
For the massive sample, the median offset is $-0.142$ dex, with an NMAD of 0.179 dex, indicating a larger systematic difference than for the full sample.

We further examine how these alternative assumptions affect the selection of the most massive galaxies. As summarized in Table~\ref{tab:robustness}, excluding the intentionally simplified constant-SFH model, $\sim$77-90\% of the galaxies in our final sample remain above the adopted evolving stellar-mass threshold under the different fitting assumptions. In particular, among the 61 massive galaxies with available MIRI photometry, 55 (90\%) remain above the mass threshold after including the MIRI data. 
We therefore conclude that, although the inferred stellar masses of individual galaxies vary modestly under different assumptions, the selection of the 200 most massive galaxies, and hence the scientific conclusions presented throughout this paper, remain robust.

In addition, we assess how the same alternative SED modeling assumptions affect SFR$_{\rm 100\,Myr}$ and SFR$_{\rm 10\,Myr}$ for both the full spectroscopic sample and the 200 massive galaxies. The comparisons and discussion are presented in Appendix~\ref{appendix3}.

\begin{table*}
\caption{Median physical properties of prism galaxies and their 16th-84th percentile ranges.}   
\centering
\renewcommand{\arraystretch}{1.2} 
\begin{threeparttable} 
 
\begin{tabular}{l c c c c c}    
\hline\hline       
Properties  & Total & Massive sample & Massive normal SFGs & Massive dusty SFGs & Massive QGs\\ 
\hline  
$N_{\rm gals}$ &6312 &200&105&66&29\\
$z_{\rm spec}$ & $4.38_{-1.01}^{+1.64}$ & $4.60_{-1.14}^{+4.68}$ & $7.51_{-3.12}^{+3.06}$ & $3.69_{-0.47}^{+1.03}$ & $3.97_{-0.75}^{+0.66}$ \\
log$(M_{\star}/M_\odot)$& $8.41_{-0.84}^{+0.80}$ &
$10.42_{-1.35}^{+0.40}$ &
$9.34_{-0.59}^{+1.09}$ &
$10.72_{-0.24}^{+0.19}$ &
$10.70_{-0.29}^{+0.24}$    \\
log(sSFR$_{100}$/yr$^{-1}$) & $-8.14_{-0.35}^{+0.16}$ & $-8.39_{-1.00}^{+0.34}$ & $-8.22_{-0.48}^{+0.17}$ & $-8.60_{-0.42}^{+0.61}$ & $-10.35_{-1.13}^{+0.67}$ \\
log(sSFR$_{10}$/yr$^{-1}$) & $-8.07_{-0.59}^{+0.69}$ & $-8.38_{-0.80}^{+0.46}$ & $-8.13_{-0.52}^{+0.31}$ & $-8.53_{-0.38}^{+0.44}$ & $-10.03_{-1.27}^{+0.34}$ \\
$A_{\rm V}$ [mag] & $0.12_{-0.09}^{+0.19}$ & $0.59_{-0.45}^{+1.47}$ & $0.30_{-0.21}^{+0.38}$ & $2.04_{-0.52}^{+0.64}$ & $0.55_{-0.30}^{+0.20}$ \\
\hline
\end{tabular}
\label{table1} 
\end{threeparttable} 
\end{table*}
\begin{figure*}
\centering
\includegraphics[width=8cm]{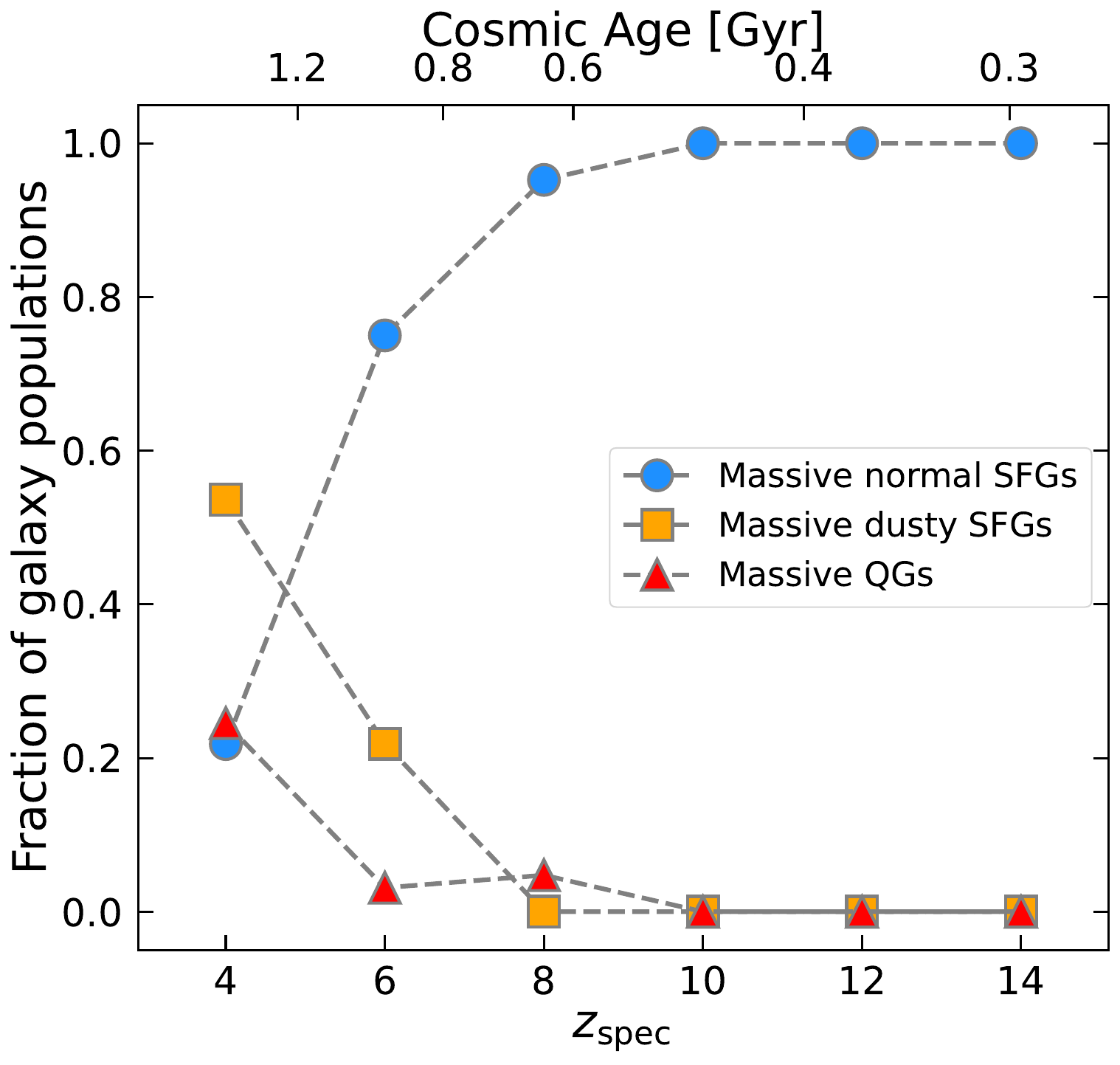}
\includegraphics[width=8cm]{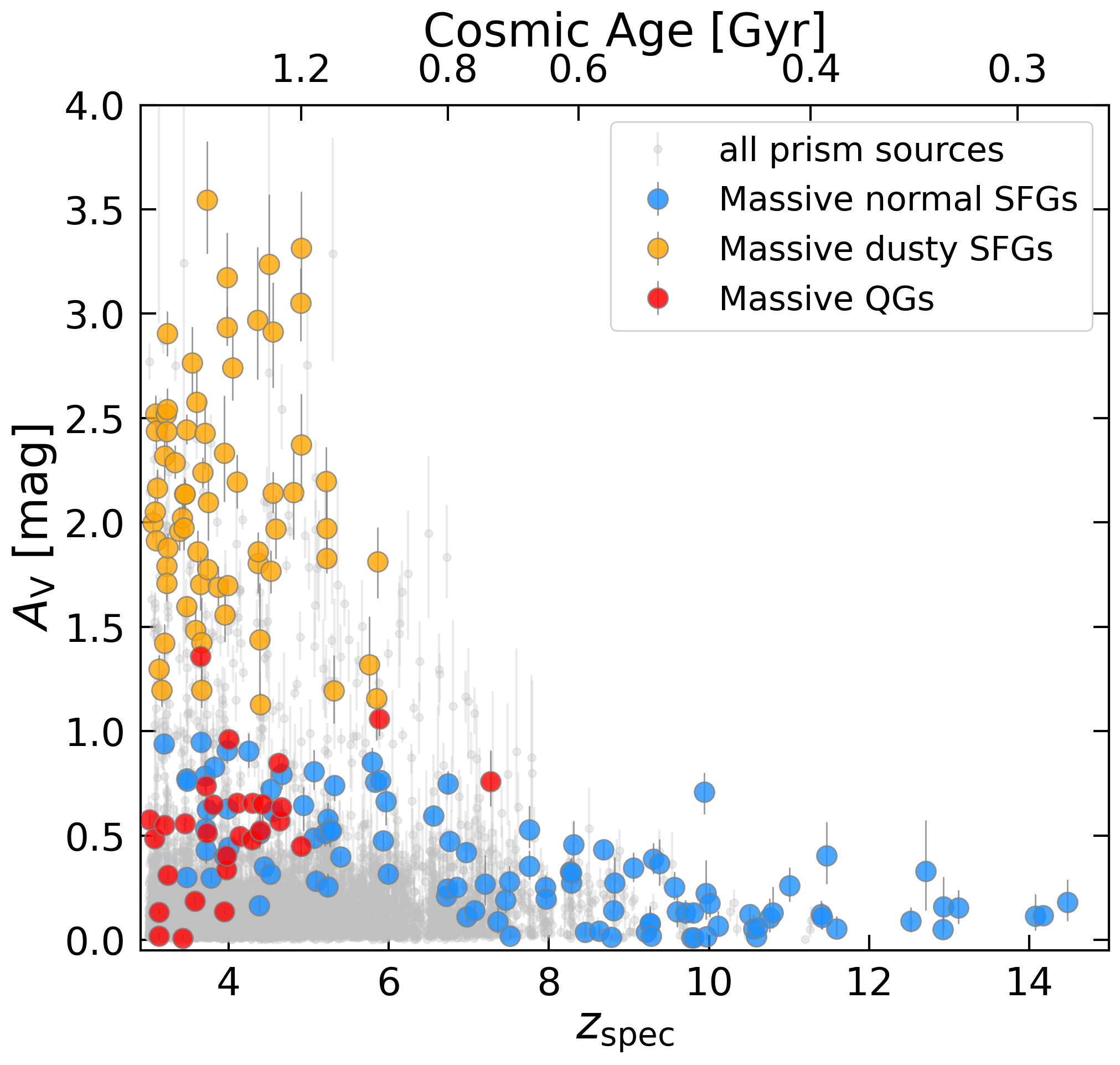}
\caption{\textbf{Evolution of massive galaxy populations and dust attenuation with redshift.}
\textit{Left:} Fraction of massive QGs, dusty SFGs, and normal SFGs as a function of $z_{\rm spec}$. We caution that since our sample is compiled from multiple JWST programs with different selection functions, these fractions should not be interpreted as the intrinsic population fractions, but rather as indicative of the overall evolutionary trends with redshift. 
\textit{Right:} $A_V$ as a function of $z_{\rm spec}$ for all prism sources (grey points) and massive galaxy populations. Error bars indicate the 16th-84th percentile range. It reveals a systematic increase in dust attenuation and the emergence of dusty SFGs toward lower redshift.
}
\label{dust}
\end{figure*}

\section{Classification of massive galaxies}\label{Sect:clean massive}

Given the clean sample of 200 most massive galaxies, we classify them into three broad categories according to their recent star-formation activity and dust content: 1) quiescent galaxies (QGs); 2) dust-obscured star-forming galaxies (dusty SFGs); and 3) normal star-forming galaxies (normal SFGs).

To select QGs, we apply an evolving specific star-formation rate (sSFR) criterion \citep[criterion \#1; e.g.,][]{Carnall2024, Gallazzi2014, Xiao2026_dormant}, 
\begin{equation}\label{eq:sSFR}
   \mathrm{sSFR}_{\rm 100\,Myr} < 0.2/t_{\rm H}(z),
\end{equation}
where $t_{\rm H}(z)$ is the cosmic age at each galaxy’s redshift. Since sSFR$_{\rm 100Myr}$ averages over the last 100 Myr, we find that some recently quenched or “dormant’’ systems with strong Balmer breaks and negligible current star formation are not captured by the above criterion \#1 alone. To recover these galaxies, we add an additional criterion \#2 based on spectral features: the strength of the Balmer break and the H$\alpha$ equivalent width (EW), similar to \cite{Covelo-Paz2025}:
\begin{align}\label{eq:dormant}
F_{\nu,4200}/F_{\nu,3500} > 1.4 \quad \& \quad A_V < 1 \text{mag}, \\
\mathrm{EW}_{\mathrm{H}\alpha}^\mathrm{rest} < 50\,\text{\AA}.
\end{align}
We calculate $F_{\nu,4200}/F_{\nu,3500}$ and $\mathrm{EW}_{\mathrm{H}\alpha}^\mathrm{rest}$ directly by fitting the observed spectra.   
The Balmer break strength ($F_{\nu,4200}/F_{\nu,3500}$) is calculated as the ratio of the $f_\nu$ values between two sides of the Balmer break, a blue window from $3400-3600\AA$ and a red window from $4150-4350\AA$. The $\mathrm{EW}_{\mathrm{H}\alpha}^\mathrm{rest}$ is the rest-frame H$\alpha$ EW calculated by dividing the \ha\, flux by the \ha\, continuum. Given the limited prism resolution, \ha\, is blended with \nii~doublet, we therefore apply a small correction, assuming that 90\% of the blended flux stems from the \ha\, line. Intriguingly, we found all galaxies selected as dormant by this criterion \#2 show very weak [O\,\textsc{iii}]~$\lambda5008$ emission, with $\mathrm{EW}_{\mathrm{[O\,\textsc{iii}]}5008}^\mathrm{rest}< 40$ \AA, further supporting their quiescent nature. 

In total, we identify 29 QGs, of which 5 satisfy only the dormant selection (criterion  \#2) and 16 satisfy both the sSFR and dormant criteria (criterion  \#1 \& criterion \#2).

The remaining galaxies are classified as star-forming and further divided into dusty SFGs and normal SFGs based on their dust attenuation derived from the joint SED fits. We classify them into dusty SFGs and normal SFGs depending on the $A_V > 1$ mag or $A_V < 1$ mag, respectively \citep[e.g.,][]{Barrufet2025}\footnote{We note that, in this work, dusty and normal SFGs are distinguished solely by $A_V = 1$ mag. Dusty SFGs may include, but are not equivalent to, so-called SMGs or DSFGs selected via submillimeter observations, while normal SFGs ($A_V < 1$ mag) may still contain some dust.}
. 

Finally, the breakdown of our sample is: 29 QGs (including 5 recently quenched dormant galaxies that satisfy only criterion  \#2), 66 dusty SFGs, and 105 normal SFGs. We also test the classification of these 200 massive galaxies under alternative SED modeling assumptions. Details are presented in Appendix~\ref{appendix_classification}.

We note that our sample is not volume-complete due to multiple JWST programs (Table~\ref{surveys}) with different target selections and depths. Therefore, these numbers do not exactly represent intrinsic fractions. However, we still expect that our galaxy sample is broadly representative of the underlying galaxy population.

\section{Results}\label{Sec: result}

\subsection{Distribution of massive galaxy populations} \label{Sec: demographics}

Figure~\ref{fig2} shows the stellar mass vs. $z_{\rm spec}$ distribution of the 200 most massive galaxies identified from the full sample of 6,312 spectroscopically confirmed galaxies at $z_{\rm spec}>3$. By construction, the stellar-mass threshold decreases toward higher redshift following the evolving halo-mass function (Sect.~\ref{Sect:selection}), allowing us to trace a representative population of the most massive galaxies over the first two billion years of cosmic history. The sample spans a wide redshift range of $z\sim3-15$, covering the full range of massive galaxies identified in the current public JWST/NIRSpec prism sample.

Figure~\ref{dust} illustrates how the composition of the massive galaxy population evolves with cosmic time. At $z\sim3-5$, dusty SFGs constitute a substantial fraction of the massive galaxy population. Toward higher redshift, the fraction of normal SFGs rises rapidly and becomes dominant by $z\gtrsim6$, while dusty SFGs and QGs become increasingly rare. This transition is also reflected in the dust properties of the massive galaxies. Most galaxies at $z\gtrsim6$ have $A_V<1$ mag, whereas heavily obscured systems are primarily found at lower redshift. Considering that our sample is compiled from programs with different selection functions, we also test whether these trends hold within individual programs (CAPERS, GTO-WIDE, JADES, MoM, and RUBIES). We find that the trends in population fractions and $A_V$ with redshift are generally consistent across programs and with Fig.~\ref{dust}.

The decline of dust attenuation toward higher redshift is consistent with a picture in which the earliest massive galaxies formed before substantial metal enrichment and dust production had occurred. The lower stellar masses (and likely lower metallicities) of the high-redshift galaxies may also contribute to this trend \citep[e.g.,][]{Casey2026, Faisst2026}. 
Together, these trends suggest that substantial dust enrichment and the emergence of quenched massive systems become common only after the first $\sim1$ Gyr of cosmic evolution.

The median physical properties of each population are summarized in Table~\ref{table1}. Massive dusty SFGs and QGs have comparable stellar masses ($\log M_\star/M_\odot \sim 10.7$), while massive normal SFGs extend to lower masses due to their broader redshift distribution and the evolving mass selection. By construction, dusty SFGs exhibit substantially higher dust attenuation ($A_V=2.04_{-0.52}^{+0.64}$ mag) than normal SFGs ($A_V=0.30_{-0.21}^{+0.38}$ mag), whereas QGs ($A_V=0.55_{-0.30}^{+0.20}$ mag) show suppressed star formation without extreme obscuration.

\begin{figure*}
\centering
\includegraphics[width=8.5cm]{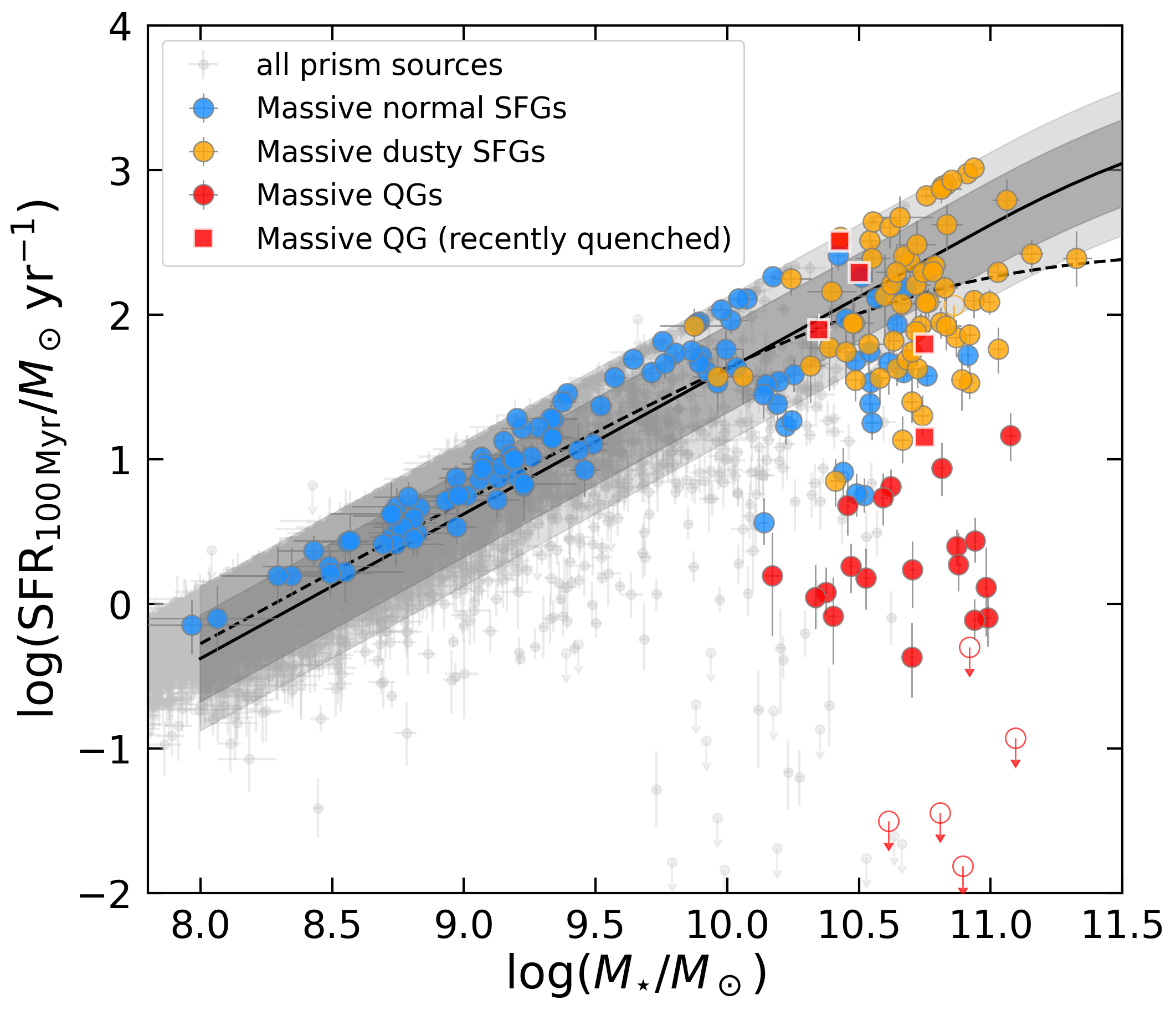}\hspace{-1mm}
\includegraphics[width=8.5cm]{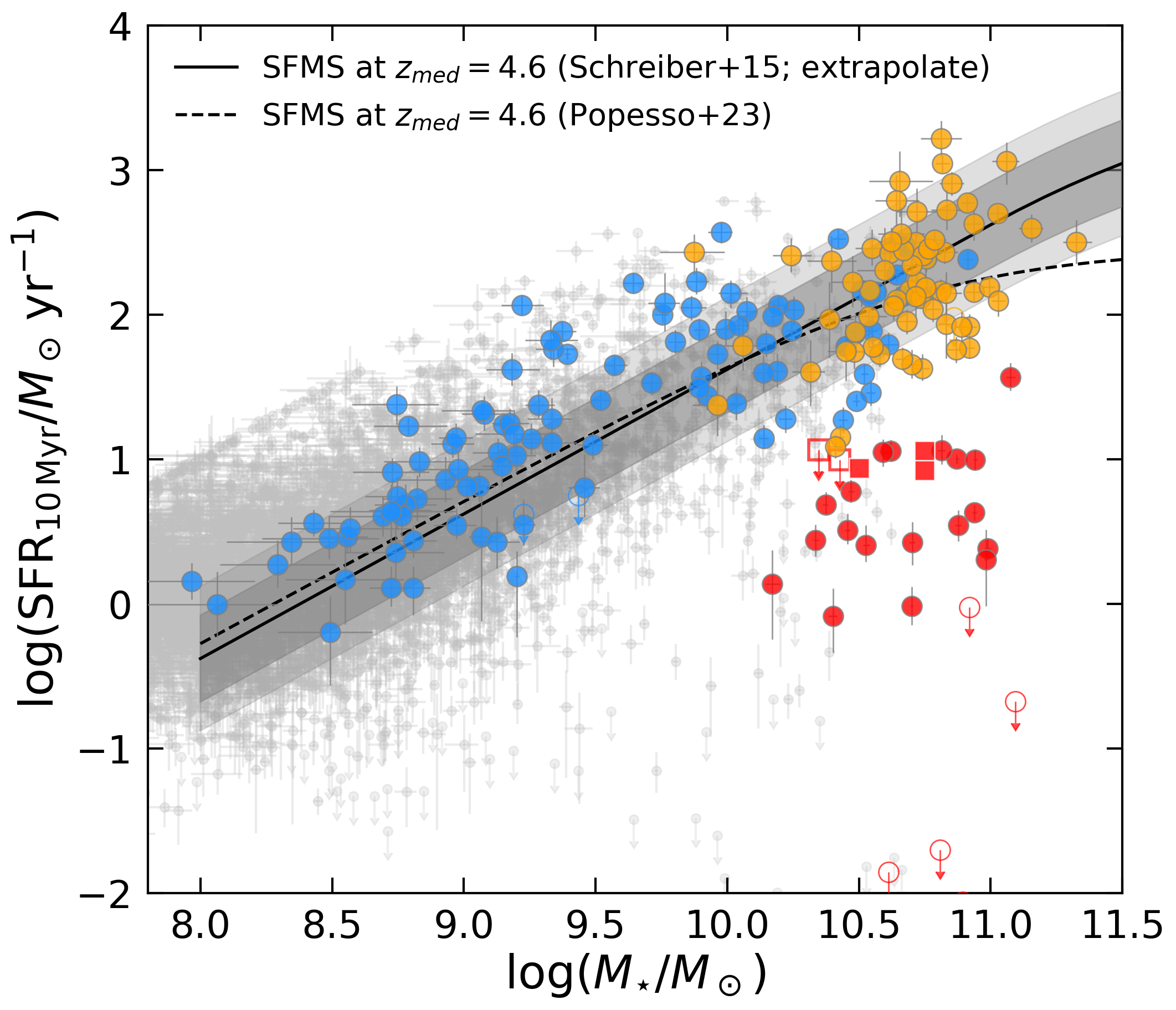}\hspace{-1mm}
\caption{\textbf{Location of massive QGs, dusty SFGs, and normal SFGs, together with all prism sources (grey), in the SFR-$M_{\star}$ plane.} The \textit{left} panel shows SFR averaged over the last 100 Myr, while the \textit{right} panel shows SFR averaged over the last 10 Myr. Massive QGs, dusty SFGs, and normal SFGs are shown in red, orange, and blue, respectively. The five recently quenched dormant galaxies are highlighted with red-filled squares. 
Error bars correspond to $1\sigma$ uncertainties. 
The \cite{Schreiber2015} SFMS at $z=4.6$ (medium redshift of the massive sample), 1$\sigma$ scatter ($\sim$0.3 dex), and a more extended typical scatter of $0.5$ dex are highlighted with a black solid line, a dark grey shaded area, and a light grey shaded area, respectively. 
The \cite{Popesso2023} SFMS at $z=4.6$ is also plotted as a black dashed line. Unfilled points show the 1$\sigma$ upper limits.
}
\label{ms}
\end{figure*}
\subsection{Star-formation activity and location relative to the star-forming main sequence} \label{Sec: properties}

Figure~\ref{ms} places the massive galaxies relative to the star-forming main sequence (SFMS). When adopting SFR$_{\rm 100\,Myr}$, i.e., the SFR averaged over the last 100 Myr (left panel), massive normal SFGs and dusty SFGs lie predominantly on the SFMS, while QGs are typically $\gtrsim2$ dex below it. When using SFR$_{\rm 10\,Myr}$ (right panel), actively star-forming systems exhibit a larger scatter but remain broadly consistent with the SFMS. The increased dispersion is consistent with a stochastic or bursty mode of star formation \citep[e.g.,][]{Ciesla2024, Simmonds2025, McClymont2025}.

Notably, the five recently quenched (dormant) galaxies lie on the SFMS when using SFR$_{\rm 100\,Myr}$, indicating substantial star formation within the past 100 Myr. In contrast, when adopting SFR$_{\rm 10\,Myr}$, these galaxies fall $\sim$1-2 dex below the SFMS. This divergence between SFR$_{\rm 100\,Myr}$ and SFR$_{\rm 10\,Myr}$ directly indicates a rapid recent decline in star formation, supporting their interpretation as recently and rapidly quenched systems. The existence of such galaxies suggests that at least some massive galaxies transition from active star formation to quiescence on timescales significantly shorter than 100 Myr.

In addition, we note that none of the massive galaxies lie far above the SFMS in an extreme starburst regime in the left panel. This apparent upper boundary is primarily driven by the adopted SFR timescale. Since SFR$_{\rm 100\,Myr}$ is averaged over 100 Myr, short-lived starbursts are smoothed out, lowering its apparent SFR at fixed stellar mass.  A similar but weaker effect is present when using SFR$_{\rm 10\,Myr}$, where the shorter averaging timescale allows for a somewhat larger scatter toward high SFRs. 
Therefore, the lack of extreme outliers above the SFMS does not imply a physical cutoff in star formation, but instead reflects the smoothing effect introduced by the chosen SFR timescale.

Overall, the majority of massive galaxies at $z\sim3-15$ are broadly consistent with the SFMS, while a substantial population of quenched and recently quenched systems is already in place within the first two billion years of cosmic time. Together with the population trends presented in Sect.~\ref{Sec: demographics}, this suggests that the emergence of massive quiescent galaxies is accompanied by a progressive build-up of dust-rich and rapidly evolving massive systems at early cosmic epochs.

\begin{figure*}
\centering
\includegraphics[width=8cm]{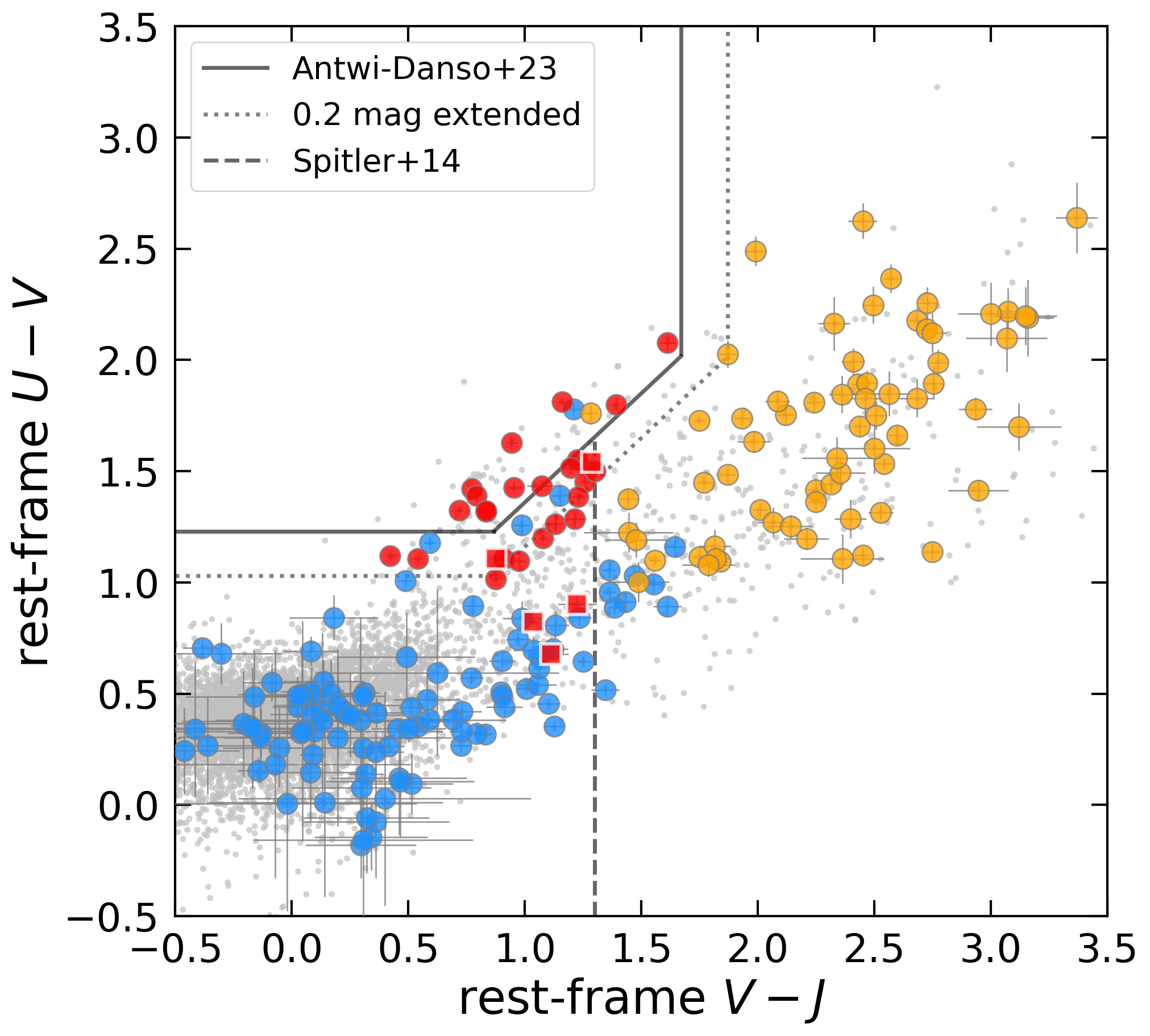} 
\includegraphics[width=8cm]{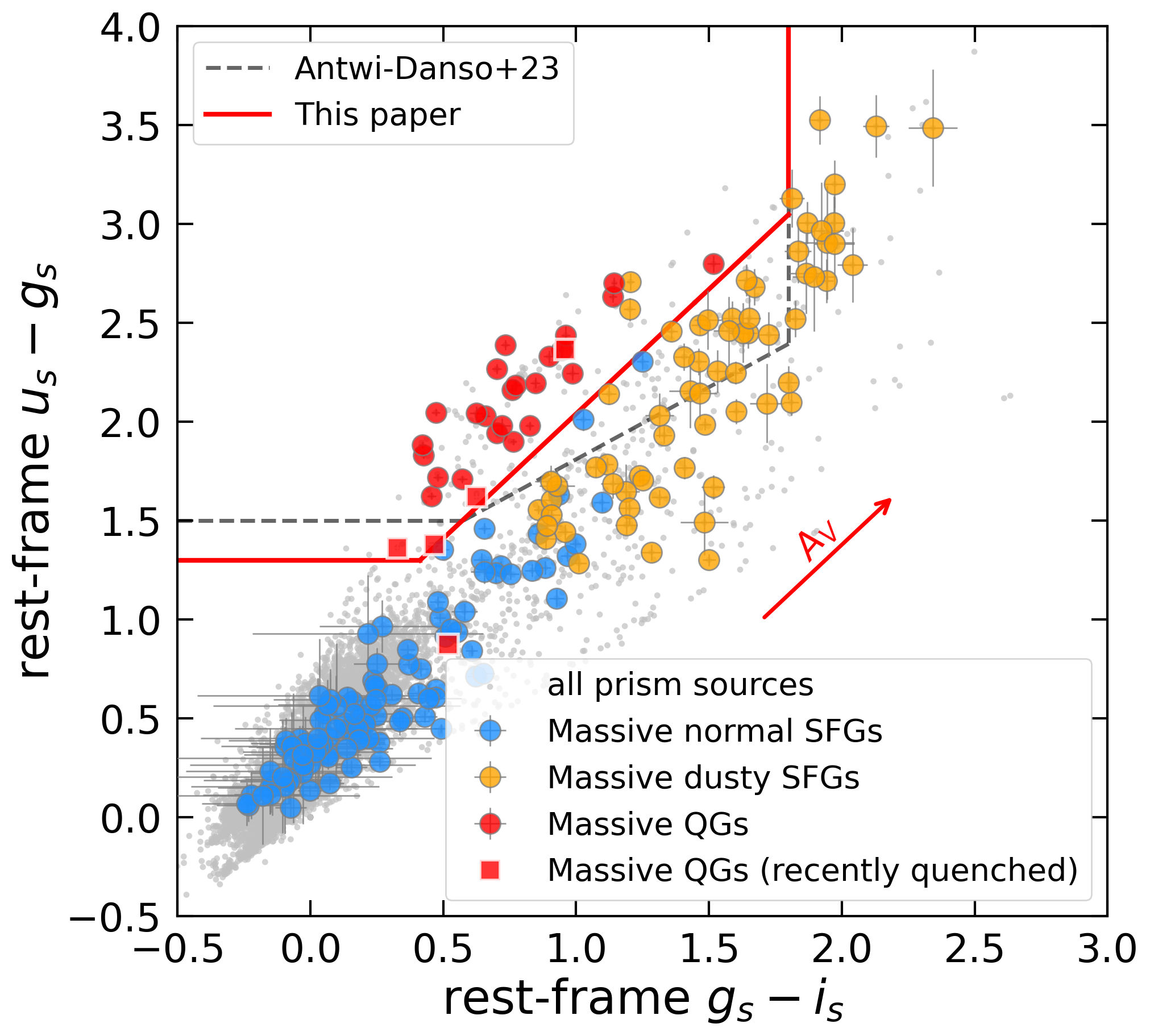}
\caption{\textbf{Comparison between rest-frame \textit{UVJ} (\textit{left}) and (\textit{ugi})$_{\rm s}$ (\textit{right}) selections and our spectroscopically classified massive galaxies. }
Grey points show all prism galaxies at $z_{\rm spec}>3$, while the 200 most massive galaxies are overplotted and color-coded as massive QGs (red), dusty SFGs (orange), and normal SFGs (blue). 
In the \textit{left} panel, the black lines indicate the classical \textit{UVJ} quiescent selection \citep{Antwi-Danso2023}, with the dotted extension showing a relaxed boundary allowing an additional 0.2 mag padding. The dashed line is used to separate the normal SFGs and dusty SFGs \citep{Spitler2014}. 
In the \textit{right} panel, the black dashed line shows the (\textit{ugi})$_{\rm s}$ quiescent criterion proposed by \cite{Antwi-Danso2023}, while the red line indicates the empirical boundary introduced in this work (Eq.~\ref{ugi}), optimized to separate better spectroscopically identified massive QGs from dusty SFGs and normal SFGs. The red arrow indicates the reddening vector.
}
\label{uvj_ugis}
\end{figure*}

\subsection{\textit{UVJ} and (\textit{ugi})$_{\rm s}$}\label{Sec:uvj}

Rest-frame color-color selections such as \textit{UVJ} have long provided a convenient, purely photometric separation between quiescent and star-forming galaxies \citep[e.g.,][]{Williams2009, Brammer2011, Spitler2014, muzzin2013b, Valentino2023, Belli2019, Jizhiyuan2026, Baker2025_FLAMINGO}. 
In Fig.~\ref{uvj_ugis}, we compare these widely used selection criteria to our sample of 200 most massive galaxies, whose classifications are based on sSFR and spectroscopic diagnostics (Sect.~\ref{Sect:clean massive}). In the \textit{UVJ} plane (Fig.~\ref{uvj_ugis}-\textit{left}), a clear mismatch shows: while many of our massive QGs lie within the classical quiescent box, a fraction of $>$30\% massive QGs -- especially the recently quenched systems -- fall close to the boundary or outside the \textit{UVJ} box, even when adopting a looser criterion allowing for an extra padding of 0.2 mag. At the same time, several massive normal SFGs and dusty SFGs are also scattered into the nominal quiescent region.

This mismatch is not unexpected in the JWST era and highlights that \textit{UVJ} boundaries calibrated primarily at lower redshift are not necessarily portable to massive galaxies at $z\gtrsim3$ \citep[e.g.,][]{Antwi-Danso2023, Gould2023, Long2024, Whitaker2026_REVIEW}. One issue is: high-$z$ QGs typically host younger stellar populations, resulting in relatively shallow Balmer breaks and therefore bluer $(U-V)$ colors compared to lower-redshift quiescent systems. As a result, recently quenched massive galaxies can fail the traditional \textit{UVJ} selection \citep[e.g.,][]{Park2023}. Second, \textit{UVJ} relies critically on a robust estimate of rest-frame $J$; however, at $z>3$ the rest-frame $J$ band has already moved beyond the longest JWST/NIRCam coverage (F444W), making $V-J$ color increasingly sensitive to extrapolation and SED-template assumptions. Such extrapolation can alter rest-frame $(V-J)$ by up to $\sim$1 mag, directly affecting whether a galaxy falls inside or outside the \textit{UVJ}-quiescent region \citep{Antwi-Danso2023}. Third, strong rest-frame optical emission lines and dust attenuation can bias broad-band fluxes and mimic redder $(U-V)$ colors, making some star-forming galaxies appear quiescent in \textit{UVJ} space. This contamination becomes increasingly important at high redshift as extreme emission-line galaxies become more common.

Motivated by these limitations, \cite{Antwi-Danso2023} proposed the synthetic (\textit{ugi})$_{\rm s}$ diagram, designed to reduce sensitivity to rest-frame $J$ extrapolation and emission-line contamination. In Fig.~\ref{uvj_ugis}-\textit{right}, the (\textit{ugi})$_{\rm s}$ space indeed produces a tighter sequence than \textit{UVJ}. However, when adopting the original \cite{Antwi-Danso2023} quiescent boundary (dashed line), we find that a significant number of our massive dusty SFGs fall inside the nominal QG region. In other words, for our massive sample, this criterion encloses not only spectroscopically confirmed QGs but also a large fraction of dusty SFGs. A similar issue has also been reported by \cite{ZhangYunchong2026}. This contamination directly motivates us to redefine the boundary.

We therefore introduce an empirical (\textit{ugi})$_{\rm s}$ criterion tailored to our spectroscopic massive sample (red line in Fig.~\ref{uvj_ugis}-\textit{right}):
\begin{equation}\label{ugi}
  \begin{split}
  ~~  (u_s - g_s) ~ > ~ &1.3,  \\
  ~~  (g_s - i_s) ~ < ~ &1.8,  \\
  ~~  (u_s - g_s) ~ > ~ &(g_s - i_s) \times 1.26 + 0.78.
    \end{split}
\end{equation}
The new boundary is adjusted to better isolate spectroscopically selected massive QGs, while minimizing overlap with the massive dusty SFGs and massive normal SFGs.

In addition, given that these color criteria are commonly used for the photometry-only samples, we also tested the location of our sample using the rest-frame colors derived from photometry-only SED fitting (more details see Fig.~\ref{uvj_ugis_phot}). Although individual galaxies shift in color space, our new (\textit{ugi})$_{\rm s}$ criterion still reduces the overlap between QGs and dusty SFGs.

Note that even in (\textit{ugi})$_{\rm s}$ space, rest-frame colors remain sensitive to SED modeling and limited long-wavelength coverage. In particular, the rest-frame \textit{i} band shifts beyond F444W at $z>6$, making the inferred colors increasingly dependent on extrapolation. In addition, small shifts within the color uncertainties can move galaxies across the selection boundaries. As a result, color-color selections can help separate high-$z$ massive QGs, dusty SFGs, and normal SFGs in a statistical sense, but contamination between the populations cannot be fully avoided. Throughout this work, we therefore adopt spectroscopically anchored criteria (sSFR$_{\rm 100\,Myr}$, Balmer-break strength, and $\mathrm{EW}_{\mathrm{H}\alpha}^\mathrm{rest}$) as our primary classification scheme.

\begin{figure*}
\centering
\includegraphics[width=18.5cm]{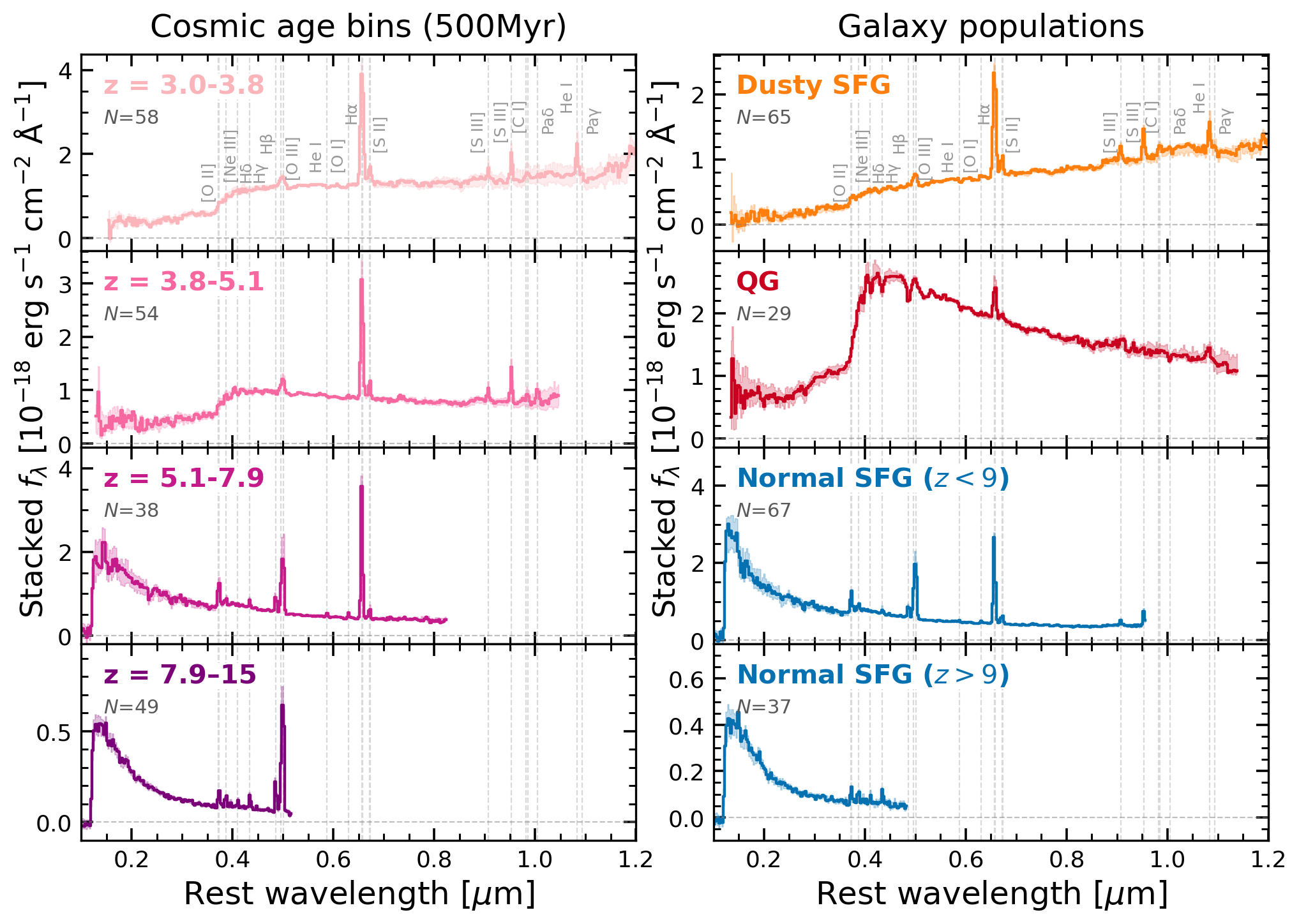}
\caption{\textbf{Stacked spectra of massive galaxies in redshift bins (\textit{left}; 500 Myr intervals) and by galaxy population (\textit{right}).}
All spectra are shifted to the rest-frame and median-stacked after slit-loss correction. Shaded regions indicate the $16$th-$84$th percentile range from bootstrap resampling (see Sect.~\ref{Sec: stacking}). Vertical grey dashed lines indicate prominent emission/absorption features. The number of contributing galaxies at each wavelength is shown in each panel.
Normal SFGs are split into $z<9$ and $z>9$ bins to account for their broad redshift range and the different underlying stellar populations. At $z>9$, stellar-mass estimates become increasingly uncertain due to the lack of photometric coverage redward of F444W.
}
\label{fig:stack}
\end{figure*}

\begin{table*}
\caption{Physical properties of median stacked massive galaxies in redshift bins and by galaxy population.}   
\centering
\renewcommand{\arraystretch}{1.2} 
\begin{threeparttable} 
 
\begin{tabular}{l c c c c c c}    
\hline\hline       
                    
Stack parameter  & $N_{\rm gals}$ & $z_{\rm spec,med}$ & log$(M_{\star}/M_\odot)$ & log(SFR$_{100}$/$M_{\odot}$yr$^{-1}$)  & $A_{\rm V}$ [mag] & $t_{\rm form}$ [Gyr]\\ 
\hline  
\multicolumn{6}{l}{$z_{\rm spec}$ (500 Myr cosmic age bin)}\\
\hline  
$[3.0\text{-}3.8]$ & 58 & 3.44 & $10.67_{-0.02}^{+0.02}$ & $2.40_{-0.10}^{+0.08}$ & $2.17_{-0.09}^{+0.06}$ & $1.63_{-0.08}^{+0.04}$ \\
$[3.8\text{-}5.1]$ & 54 & 4.38 & $10.43_{-0.02}^{+0.02}$ & $1.87_{-0.11}^{+0.09}$ & $1.34_{-0.06}^{+0.06}$ & $1.12_{-0.04}^{+0.03}$ \\
$[5.1\text{-}7.9]$ & 38 & 5.95 & $9.95_{-0.03}^{+0.04}$ & $2.00_{-0.03}^{+0.03}$ & $1.32_{-0.05}^{+0.06}$ & $0.90_{-0.04}^{+0.01}$ \\
$[7.9\text{-}15]$  & 49 & 9.81 & $8.87_{-0.05}^{+0.05}$ & $0.60_{-0.07}^{+0.07}$ & $0.10_{-0.02}^{+0.04}$ & $0.32_{-0.02}^{+0.03}$ \\
\hline
\multicolumn{6}{l}{galaxy population}\\
\hline  
Massive dusty SFG        & 65 & 3.69 & $10.71_{-0.02}^{+0.02}$ & $2.78_{-0.07}^{+0.04}$ & $3.13_{-0.06}^{+0.04}$ & $1.60_{-0.03}^{+0.01}$ \\
Massive QG          & 29 & 3.97 & $10.53_{-0.01}^{+0.01}$ & $0.05_{-0.30}^{+0.19}$ & $0.38_{-0.01}^{+0.01}$ & $1.12_{-0.01}^{+0.01}$ \\
Massive normal SFG ($z<9$) & 67 & 5.59 & $9.85_{-0.02}^{+0.02}$ & $1.92_{-0.04}^{+0.03}$ & $0.55_{-0.05}^{+0.04}$ & $0.93_{-0.03}^{+0.01}$ \\
Massive normal SFG ($z>9$) & 37 & 10.51 & $8.81_{-0.07}^{+0.08}$ & $0.52_{-0.11}^{+0.12}$ & $0.19_{-0.09}^{+0.18}$ & $0.29_{-0.02}^{+0.03}$ \\
\hline
\end{tabular}
\begin{tablenotes}
\item \textbf{Note.} Values correspond to physical properties derived from SED fitting of the median stacked spectra using \texttt{Bagpipes}, with uncertainties indicating the $16$th-$84$th percentile ranges. $t_{\rm form}$ is defined as the age of the Universe when 50\% of the final stellar mass had formed.
\end{tablenotes}
\label{stack} 
\end{threeparttable} 
\end{table*}

\begin{figure*}[t]
\centering
\includegraphics[width=6cm]{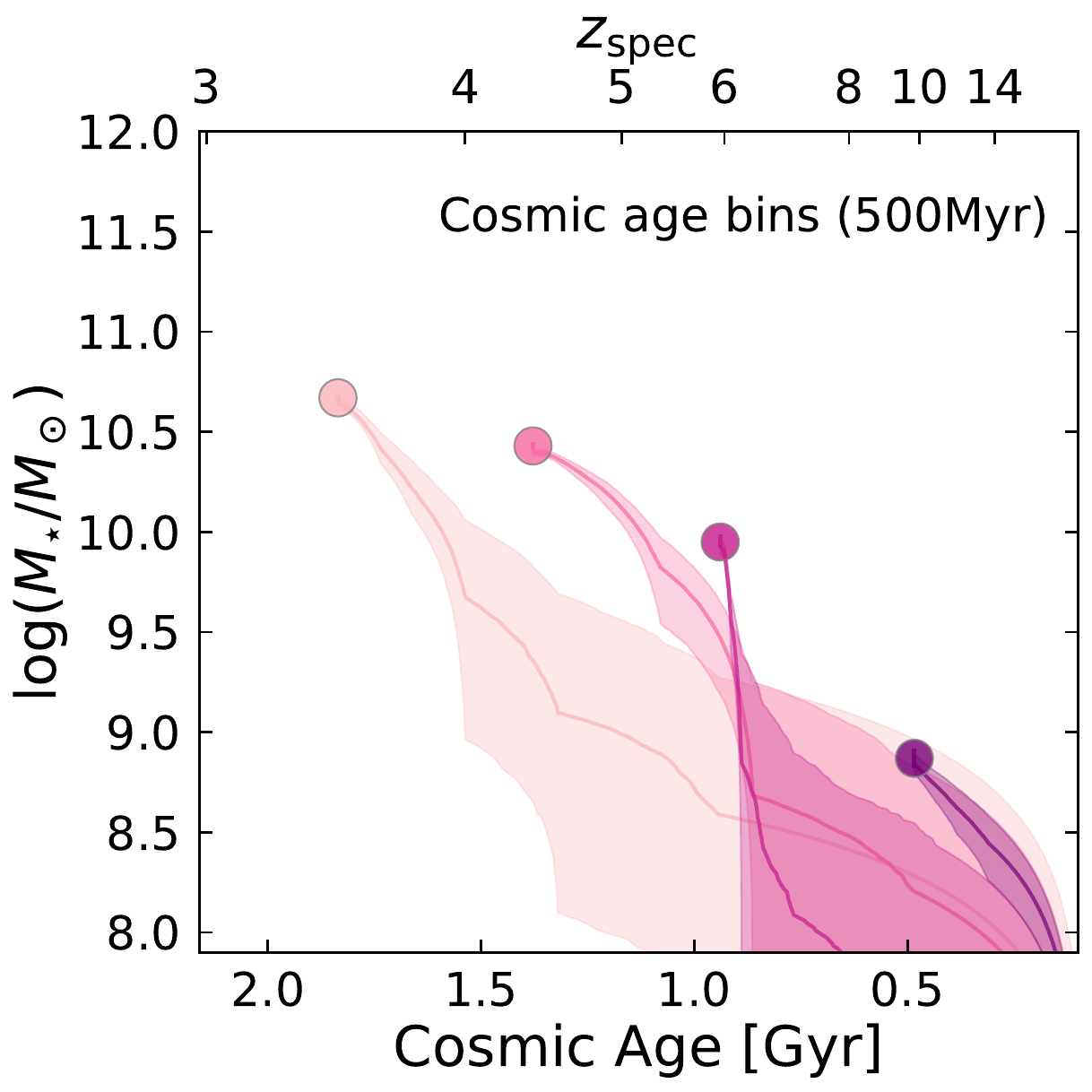} \hspace{-2mm}
\includegraphics[width=6cm]{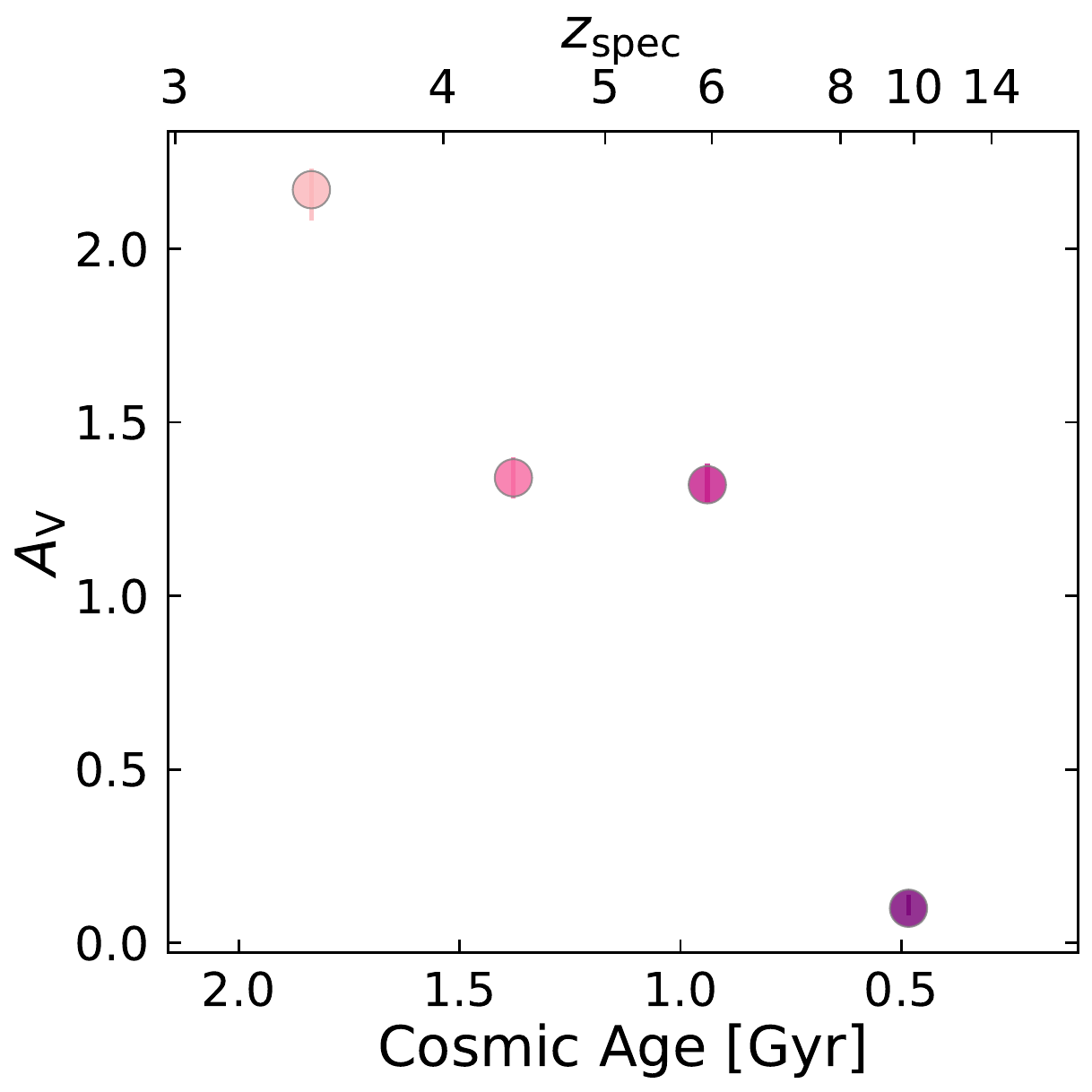}
\hspace{-2mm}
\includegraphics[width=6cm]{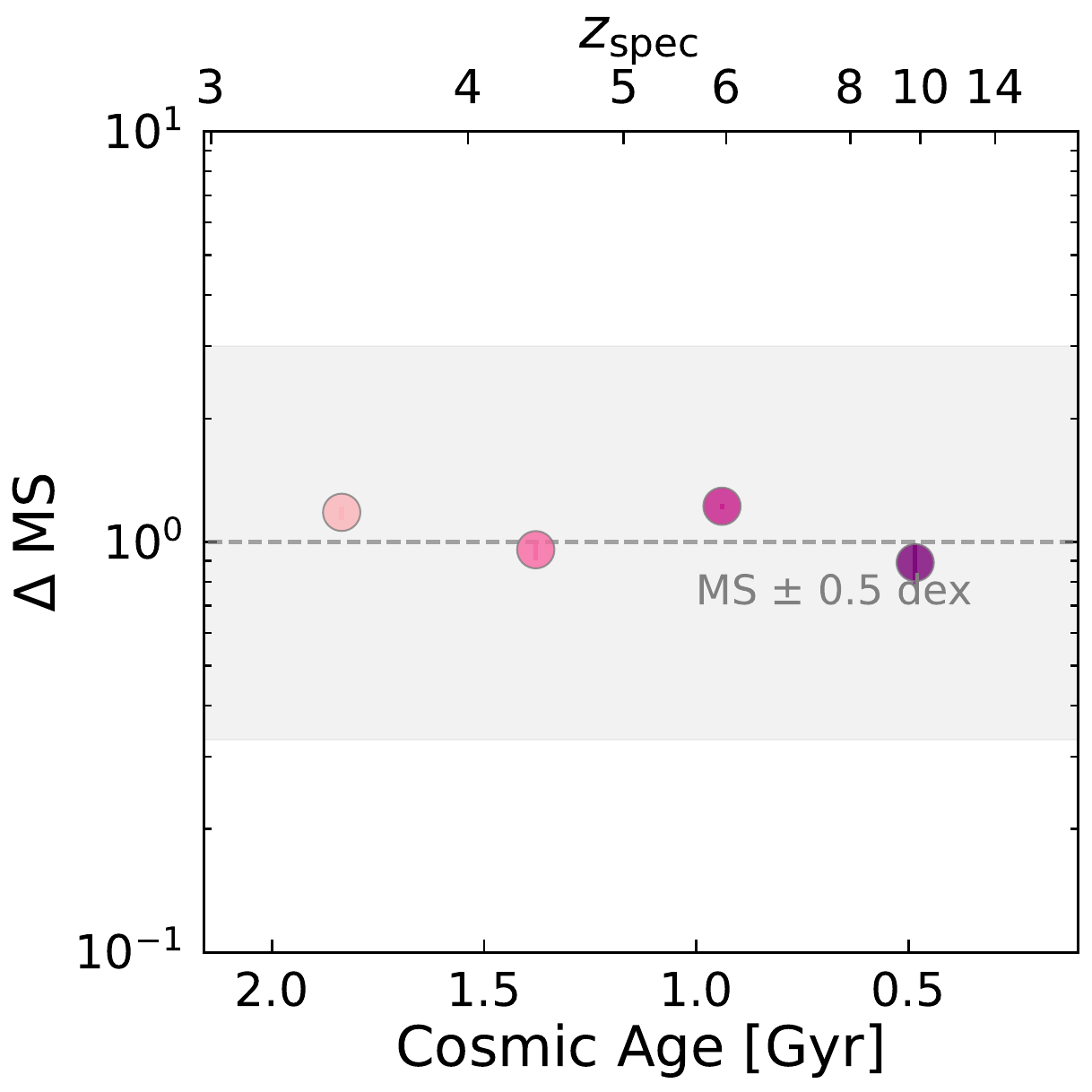}
\caption{\textbf{Evolution of median stacked physical properties as a function of cosmic age.} 
\textit{Left:} Stellar mass and stellar mass assembly histories derived from stacked SED fitting, with shaded regions indicating the $16$th-$84$th percentile range. 
\textit{Middle:} Evolution of dust attenuation ($A_V$). 
\textit{Right:} Offset from the \cite{Popesso2023} star-forming main sequence ($\Delta$MS=SFR$_{100}$/SFR$_{\rm MS}$), where the grey shaded region indicates the typical $\pm0.5$ dex scatter. 
Data points show the median values in each redshift bin with $1\sigma$ uncertainties. The top axis indicates the corresponding redshift. The stacks suggest a decrease in stellar mass and dust attenuation toward earlier cosmic times, while the massive galaxies remain broadly consistent with the main sequence within the typical scatter.
}
\label{fig:stack_trend}
\end{figure*}

 \subsection{Stacking spectra and derived properties} \label{Sec: stacking}
 To better characterize the global spectral properties of massive galaxies and to improve the S/N ratio of faint spectral features, we construct median stacked spectra for the 200 massive galaxies in redshift bins and by galaxy population (Fig.~\ref{fig:stack}). Each prism spectrum is first converted from observed flux density to rest-frame luminosity density using its $z_{\rm spec}$, thereby removing redshift-dependent effects in both wavelength and flux. Prior to stacking, slit-loss corrections are applied based on the best-fit SED of each individual galaxy (Sect.~\ref{Sect:SED}). To preserve spectral shapes while minimizing luminosity-driven biases, each spectrum is normalized within a rest-frame window at $0.52-0.58\,\mu$m. The normalized spectra are then resampled onto a common rest-frame wavelength grid and combined using a median stacking approach. At each wavelength, stacked values are retained only if at least 10 galaxies contribute, ensuring robust statistics.
Uncertainties are estimated via bootstrap resampling, including both source resampling and perturbations based on flux uncertainties. Finally, the stacked spectra and their uncertainties are converted back to the observed frame using the median redshift of each bin. 

The stacked spectra in redshift bins (Fig.~\ref{fig:stack}-\textit{left}) reveal a clear evolution with cosmic time. At the highest redshifts ($z \gtrsim 8$), the spectra are dominated by blue continua with weak Balmer breaks, indicating young stellar populations and low dust attenuation. Toward lower redshift, the continuum becomes progressively redder, with the stronger Balmer break, consistent with the build-up of more evolved stellar populations. By $z \sim 3-5$, the stacked spectra show prominent Balmer breaks and red continua, indicative of significant dust attenuation and more mature stellar populations. Strong nebular emission lines (e.g., H$\alpha$, H$\beta$, and [O\,\textsc{iii}]) are detected across all redshift bins, with variations in relative strength reflecting changes in the underlying ISM conditions. These trends are consistent with the physical properties derived from SED fitting of the median stacked spectra (Table~\ref{stack}), where both stellar mass and dust attenuation decrease toward earlier cosmic times.

The stacked spectra separated by galaxy population (Fig.~\ref{fig:stack}-\textit{right}) show distinct spectral signatures. Massive normal SFGs exhibit blue continua and strong emission lines, characteristic of actively star-forming systems with modest dust attenuation. Massive dusty SFGs display significantly redder continua with suppressed UV emission, together with prominent emission lines, consistent with heavily obscured star formation. In contrast, massive QGs show pronounced Balmer breaks and weak or absent emission lines \citep[the weak feature at $\sim$6500\,\AA\ is likely dominated by blended \nii\ doublet rather than \ha; e.g., ][]{deGraaff2025}, indicating little ongoing star formation. These spectral differences provide an independent validation of our classification based on sSFR, Balmer-break strength, and dust attenuation (Sect.~\ref{Sect:clean massive}).

To further quantify these trends, we perform SED fitting on the median stacked spectra using \texttt{Bagpipes}, adopting the same parameterization and priors as for the individual galaxies (Sect.~\ref{Sect:SED}). The resulting median physical properties are summarized in Table~\ref{stack}, and their evolution with cosmic time is shown in Fig.~\ref{fig:stack_trend}. The stellar mass assembly histories (Fig.~\ref{fig:stack_trend}-\textit{left}) indicate that massive galaxies build up a substantial fraction of their stellar mass rapidly at early times, with $t_{\rm form}$ decreasing toward higher redshift in absolute time while remaining a significant fraction of the cosmic age at each epoch. 
At the same time, the dust attenuation (Fig.~\ref{fig:stack_trend}-\textit{middle}) shows a clear decline with increasing redshift, consistent with progressively lower levels of dust and metal enrichment toward earlier cosmic times.
This behavior is in good agreement with the trend observed for individual galaxies in Fig.~\ref{dust}. 
The offset from the star-forming main sequence (Fig.~\ref{fig:stack_trend}-\textit{right}) indicates that, on average, the massive galaxy population remains broadly consistent with the main sequence within the typical scatter, although individual populations exhibit significant diversity.

Taken together, the stacked spectra and their derived properties suggest a coherent evolutionary picture in which the most massive galaxies transition from predominantly blue, less obscured star-forming systems at $z \gtrsim 8$ to more dust-rich and evolved systems by $z\sim3-5$, while maintaining star-formation activity broadly consistent with the main sequence.
However, we caution that these trends may be influenced by selection effects in the current JWST spectroscopic sample, e.g., high-$z$ surveys may be biased against the dustiest galaxies at high redshift.

\subsection{Mass assembly history} \label{Sec: SFH}
\begin{figure*}[!h]
\centering
\includegraphics[width=13cm]{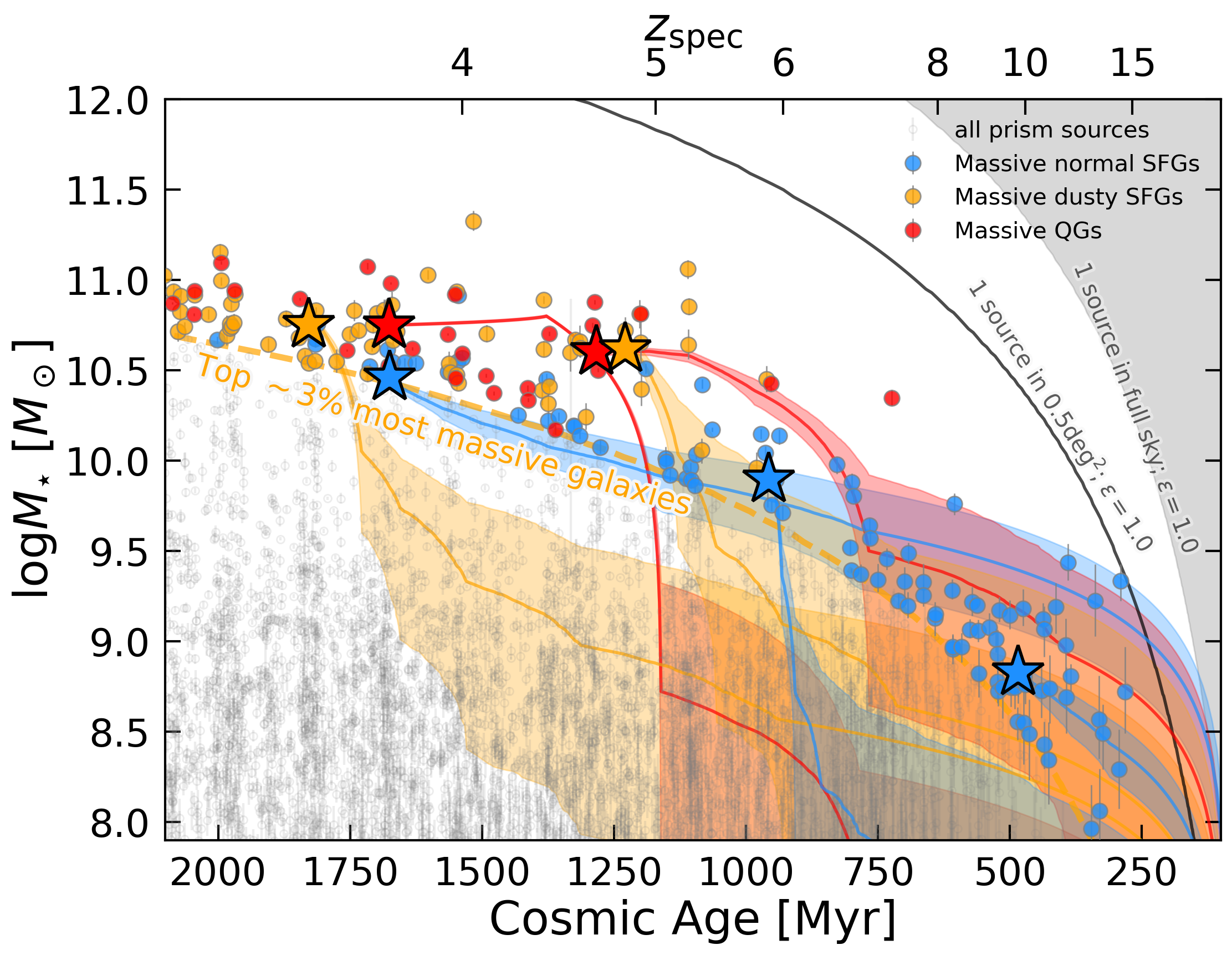} 
\caption{\textbf{Stellar mass assembly histories of massive galaxies as a function of cosmic age.}
The symbols and lines are the same as in Figure~\ref{fig2}. The stars indicate the median stellar masses of the stacked massive galaxy populations. The stacked populations are divided into three redshift intervals, $z=3-4.3$, $4.3-7.9$, and $7.9-15$, corresponding to approximately 750 Myr cosmic-age bins. The colored solid curves and shaded regions show the stellar mass assembly histories (median and 16th-84th percentile range) derived from \texttt{Bagpipes} SED fitting of the stacked spectra. 
}
\label{fig:massassembly}
\end{figure*}

The distinct spectral properties and physical characteristics of the massive QGs, dusty SFGs, and normal SFGs suggest that these populations represent different stages of massive galaxy evolution. To investigate their possible evolutionary connections, we compare the stellar mass assembly histories of the three populations derived from the median stacked spectra. Fig.~\ref{fig:massassembly} shows the resulting assembly tracks from \texttt{Bagpipes} SED fitting for the three redshift bins ($z=3.0-4.3$, $4.3-7.9$, and $7.9-15$; 750 Myr cosmic age intervals) of different massive galaxy populations. The assembly histories are overplotted on the stellar mass-cosmic age plane, along with the full prism spectroscopic sample.

The inferred assembly histories indicate that massive QGs at $z\sim4$ had already assembled most of their stellar mass within the first $\sim$1 Gyr after the Big Bang, implying that the bulk of their stellar mass was already in place roughly $\sim$500 Myr before the observed epoch and prior to quenching. This picture is supported by their stacked spectra, which exhibit pronounced Balmer breaks together with weak nebular emission lines (Fig.~\ref{fig:stack}), indicative of evolved stellar populations and strongly suppressed ongoing star formation. Such rapid early assembly is broadly consistent with recent studies suggesting that high-redshift massive quiescent galaxies often experienced bursty rather than smoothly declining SFHs \citep[e.g.,][]{Forrest2020, Nanayakkara2025}. Interestingly, the higher-redshift QG stack follows a broadly similar assembly history while reaching a comparable stellar mass at an even earlier cosmic epoch. Such an alignment suggests that the formation and quenching of massive galaxies within the first billion years of cosmic history were driven by similar mechanisms to those at lower redshifts, both of which cause rapid star formation quenching.

The massive normal SFGs exhibit markedly different assembly histories across redshift. Both the highest-redshift ($z=7.9-15$) and lowest-redshift ($z=3.0-4.3$) stacks show relatively gradual stellar-mass growth. However, the intermediate-redshift normal SFGs ($z=4.3-7.9$) display a substantially steeper assembly history, rapidly building up stellar masses of $\log(M_\star/M_\odot)\sim10$ within less than $\sim$100 Myr of cosmic time. Such rapid growth is broadly consistent with scenarios in which a large fraction of the stellar mass is assembled during a short period of intense star formation, resembling the ``monolithic'' growth picture proposed for some of the UV-bright starburst galaxies at $z\sim2-4$ \citep[e.g.,][]{Marques-Chaves2024, Dessauges-Zavadsky2025}. The diversity of assembly histories among the normal SFGs suggests that massive star-forming galaxies might not follow a single evolutionary mode, but instead experience both relatively gradual and highly accelerated phases of stellar-mass assembly.

The massive dusty SFGs show substantial stellar-mass growth at early cosmic times while maintaining intense ongoing star formation. Massive galaxies at the highest redshifts are predominantly blue systems with low dust attenuation, whereas dusty SFGs become increasingly common at later cosmic times. A natural interpretation is that the earliest massive galaxies formed during epochs when the Universe had not yet experienced sufficient metal enrichment to produce large dust reservoirs. Alternatively, dust may be more efficiently expelled by stellar feedback in lower-mass galaxies, while only the more massive systems are able to retain substantial dust reservoirs over longer timescales. Metals produced by the earlier generations of stars and SNe subsequently enriched the ISM and invoked efficient dust formation at later cosmic times. This led to the increase of dust obscuration in massive galaxies, eventually giving rise to the massive dusty SFG population commonly observed at $z\sim3-5$.

Taken together, the assembly histories reveal a diversity of growth modes among the most massive galaxies in the early Universe, ranging from relatively gradual stellar-mass build-up to extremely rapid assembly. In the following Sect.~\ref{Sec: evolution}, we will discuss the possible evolutionary connections between different massive galaxy populations and caveats.

\begin{figure*}
\centering
\includegraphics[width=15cm]{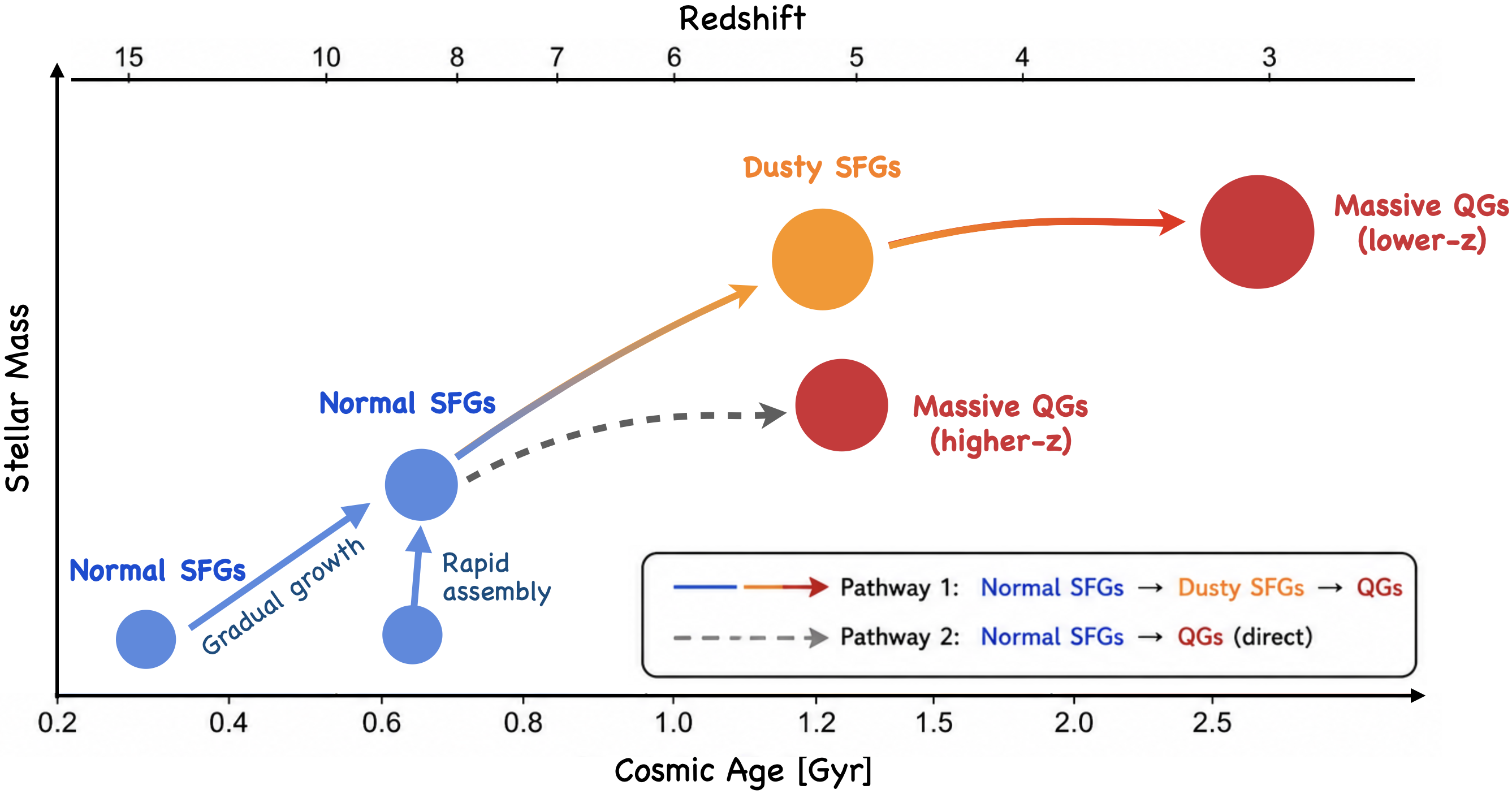} 
\caption{\textbf{Schematic illustration of two possible evolutionary pathways for the most massive galaxies in the early Universe.} Based on the average SFHs derived for the different massive galaxy populations, massive normal SFGs may first assemble through either relatively gradual growth or a rapid assembly phase. They may then evolve into dusty SFGs before quenching into massive QGs, or transition more directly into QGs without experiencing a prolonged dusty star-forming phase. The locations of the populations are illustrative and motivated by the median assembly histories shown in Fig.~\ref{fig:massassembly}. Circle sizes qualitatively represent stellar mass.}
\label{fig:cartoon}
\end{figure*}

\section{Evolutionary pathways of the most massive galaxies} \label{Sec: evolution}

The assembly histories presented in Fig.~\ref{fig:massassembly} suggest that the massive QGs, dusty SFGs, and normal SFGs may be linked through a common evolutionary framework. However, the inferred connections between these populations are not unique, and multiple evolutionary pathways may contribute to the formation of the most massive galaxies in the early Universe. The two schematic evolutionary pathways discussed below are illustrated in the conceptual diagram shown in Fig.~\ref{fig:cartoon}.

One possible pathway is that galaxies evolve from normal SFGs into dusty SFGs and subsequently cease their star formation and evolve to massive QGs \citep[e.g.,][]{Zavala2026}. This scenario is naturally supported by the observed evolution of dust attenuation. Massive galaxies at the highest redshifts are predominantly blue systems with relatively low dust attenuation and lower $M_\star$, whereas the more massive dusty SFGs become increasingly common toward lower redshift. The systematic increase in $A_V$ with cosmic time and mass suggests ongoing metal enrichment and dust production, gradually transforming initially unobscured star-forming galaxies into heavily obscured systems. In this picture, dusty SFGs represent a transitional phase between early star-forming galaxies and the massive quiescent population observed at $z\sim3-5$. The assembly histories shown in Fig.~\ref{fig:massassembly} are broadly consistent with such a sequence, where the confidence interval of massive normal SFGs at $z=7.9-15$ and $z=4.3-7.9$ (which assemble their stellar mass through both relatively gradual and rapid growth) broadly aligns with that of dusty SFGs at $z\sim5$ and massive QGs at $z\sim3-5$. 

A second pathway may involve a more direct transition from normal SFGs to QGs without passing through a prolonged dusty star-forming phase. This possibility is motivated by the highest-redshift normal SFG population ($z=7.9-15$), which already contains massive systems at very early cosmic times and may plausibly evolve into the quiescent population observed at $z\gtrsim5$, as indicated by the general consistency between their mass assembly histories in Fig.~\ref{fig:massassembly}. In this picture, quenching occurs before substantial dust enrichment has taken place, bypassing the dusty SFG phase that appears to be common among massive galaxies at later cosmic times.

Nevertheless, these evolutionary pathways should not be interpreted as direct progenitor-descendant tracks. The median assembly histories shown in Fig.~\ref{fig:massassembly} represent only the average behavior of each population, while the actual SFHs of individual galaxies are likely substantially more diverse. Furthermore, the ``most massive'' galaxies selected at different redshifts do not necessarily trace a single continuously evolving population. 
Galaxies may enter or leave the massive galaxy selection due to their diverse stellar-mass growth histories. This picture is further supported by recent cosmological simulations, which show that the most massive systems at high redshift do not necessarily evolve into the most massive systems at later epochs \citep[e.g.,][]{Baxter2026,Zhao2026}. 
As a result, the observed massive galaxy population at a given epoch likely reflects a mixture of systems caught at different stages of assembly, dust enrichment, and quenching rather than a single evolutionary sequence.

Finally, we caution that the inferred stellar mass assembly histories, particularly for dusty SFGs and normal SFGs, remain uncertain. In actively star-forming galaxies, the observed light can be dominated by young stellar populations, leading to the well-known outshining effect \citep[e.g.,][]{Narayanan2024, Gimenez-Arteaga2024}, where older underlying stellar populations become difficult to constrain. This issue is likely amplified in dusty SFGs due to heavy dust obscuration \citep[e.g.,][]{Hamed2026}. In addition, degeneracies between stellar age, dust attenuation, metallicity, and the reconstruction of complex non-parametric SFHs can broaden the allowed solutions, especially at early cosmic times. Therefore, while the relative differences between galaxy populations are likely robust, the detailed assembly timescales and early SFHs should still be interpreted with caution.

\section{Conclusions}\label{Sec: conclusion}
Using all publicly available JWST/NIRSpec prism spectroscopy in the DAWN JWST Archive (v4.5), we construct the first spectroscopic census of the 200 most massive galaxies at $z_{\rm spec}\sim3-15$. This sample represents the top 3\% most massive systems within a parent sample of 6,312 spectroscopically confirmed galaxies in JWST blank-field legacy fields, including COSMOS, UDS, EGS, GOODS-South, and GOODS-North.  By combining joint SED fitting of NIRSpec spectroscopy and multi-band photometry, we assess the robustness of their inferred stellar masses and investigate their demographics, physical properties, and evolutionary pathways of the most massive galaxies during the first two billion years of cosmic history. Our main conclusions are as follows:

\begin{enumerate}

\item We construct a clean sample of the 200 most massive galaxies after identifying and removing likely contaminants, including broad-line AGNs and LRDs (Figs.~\ref{fig1},~\ref{fig2},~\ref{fig2_}). We further demonstrate that the inferred stellar masses are robust against commonly adopted SED modeling assumptions, including different stellar population models, dust attenuation curves, star formation histories, and the inclusion of MIRI photometry (Figs.~\ref{compare_miri}, ~\ref{compare}; Table~\ref{tab:robustness}).

\item Massive galaxies exhibit a strong demographic transition with cosmic time (Fig.~\ref{dust}). At $z\gtrsim6$, the population is dominated by normal SFGs, whereas dusty SFGs and QGs become increasingly common toward lower redshift. Dust attenuation also decreases systematically toward higher redshift, indicating that the earliest massive galaxies were, on average, less chemically enriched than their lower-redshift counterparts.

\item We identify 29 massive QGs, including recently quenched galaxies whose star formation declined rapidly within the past $\sim100$ Myr (Fig.~\ref{ms}). Their locations relative to the SFMS and their strong Balmer breaks indicate that rapid quenching was already occurring within the first two billion years after the Big Bang.

\item We show that both the traditional UVJ selection and recently proposed $(ugi)_{\rm s}$ criterion exhibit significant incompleteness and contamination when applied to massive galaxies at $z>3$ (Fig.~\ref{uvj_ugis}). Motivated by our spectroscopically classified sample, we propose an updated empirical $(ugi)_{\rm s}$ criterion that provides a cleaner separation between QGs and SFGs.

\item 
Combining the stacked spectra, reconstructed SFHs, and demographic evolution (Figs.~\ref{fig:stack},~\ref{fig:stack_trend}, Table~\ref{stack}), we propose that massive galaxies likely evolve through at least two pathways toward quiescence: one proceeds through a dust-enriched phase linking normal SFGs, dusty SFGs, and QGs, while another connects normal SFGs directly to QGs without a prolonged dusty phase (Figs.~\ref{fig:massassembly},~\ref{fig:cartoon}). Massive normal SFGs themselves exhibit both relatively gradual and rapid stellar-mass assembly modes.

\item 
Together, these results suggest that rapid stellar-mass assembly, dust enrichment, and star-formation quenching were already shaping the evolutionary pathways of the most massive galaxies within the first billion years after the Big Bang.

\end{enumerate}

Our spectroscopic census provides a statistical spectroscopic
view for studying massive galaxies across the first two billion years of cosmic history and establishes an observational foundation for future comparisons with galaxy formation models and cosmological simulations. We caution that our sample is assembled from heterogeneous JWST programs (Table~\ref{surveys}) with different selection functions and is therefore not volume-complete. Nevertheless, it places important constraints on the demographics and evolutionary pathways of the earliest massive galaxies. The physical mechanisms responsible for their rapid stellar-mass assembly, dust enrichment, and quenching, however, remain poorly understood. Ongoing and future ALMA observations (e.g., Qi et al., in prep.) will provide key constraints on the cold gas and dust reservoirs of these systems, while upcoming large-area spectroscopic surveys from facilities including \textit{Euclid}, the \textit{Nancy Grace Roman Space Telescope}, and the concept \textit{AtLAST} and \textit{LST}, will extend this census to much larger cosmic volumes. Combined with cosmological simulations, these observations will enable the evolutionary pathways identified in this work to be tested quantitatively, providing a more complete picture of how the most massive galaxies assembled and evolved during the first two billion years of cosmic history.

\section*{Data availability}

The Full catalog of 200 massive galaxies at $z_{\rm spec}\sim3-15$ (Table~\ref{individual}), including the physical properties derived from joint spectrophotometric SED fitting and galaxy classifications, is available at \url{https://doi.org/10.5281/zenodo.23888775}.

\begin{acknowledgements}
We are very grateful to the anonymous referee for instructive comments, which helped improve the overall quality and strengthen the analysis of this work. 
We thank Mark Dickinson for valuable discussions and suggestions that improved this paper.
The data products presented herein were retrieved from the Dawn JWST Archive (DJA). DJA is an initiative of the Cosmic Dawn Center (DAWN), which is funded by the Danish National Research Foundation under grant DNRF140. 
This work has received funding from the Swiss State Secretariat for Education, Research and Innovation (SERI) under contract number MB22.00072, as well as from the Swiss National Science Foundation (SNSF) through project grants 200020\_207349 and 2000-1-243073. 
The Cosmic Dawn Center (DAWN) is funded by the Danish National Research Foundation under grant DNRF140.
MA is supported by FONDECYT grant number 1252054, and gratefully acknowledges support from ANID Basal Project FB210003,  ANID MILENIO NCN2024 112, and ANID + Vinculación Internacional + FOVI250261.
YF is supported by JSPS KAKENHI Grant Numbers JP23K13149 and 26K00736.
AdG acknowledges support from a Clay Fellowship awarded by the Smithsonian Astrophysical Observatory.
LB acknowledges support from the Sussex Astronomy Centre STFC Consolidated Grant 2023-2026 (ST/X001040/1).
The work presented in this paper is based on observations made with the NASA/ESA/CSA JWST. The data were obtained from the Mikulski Archive for Space Telescopes at the Space Telescope Science Institute, which is operated by the Association of Universities for Research in Astronomy, Inc., under NASA contract NAS 5-03127 for JWST.
These observations are associated with the following programmes:
\#1180, \#1181, \#1210, \#1211, \#1212, \#1213, \#1214, \#1215, \#1286, \#1287, \#1345, \#2198, \#2565, \#2750, \#3215, \#4106, \#4233, \#5224, \#5997,  \#6368, \#6585, \#8018, and \#8060.

\end{acknowledgements}

\bibliography{reference_thispaper}{}
\bibliographystyle{aa}

\onecolumn
\begin{appendix}

\section{Examples of massive galaxies. }\label{appendix_sed} 
\begin{figure*}[h!]
\centering
\includegraphics[width=17.5cm]{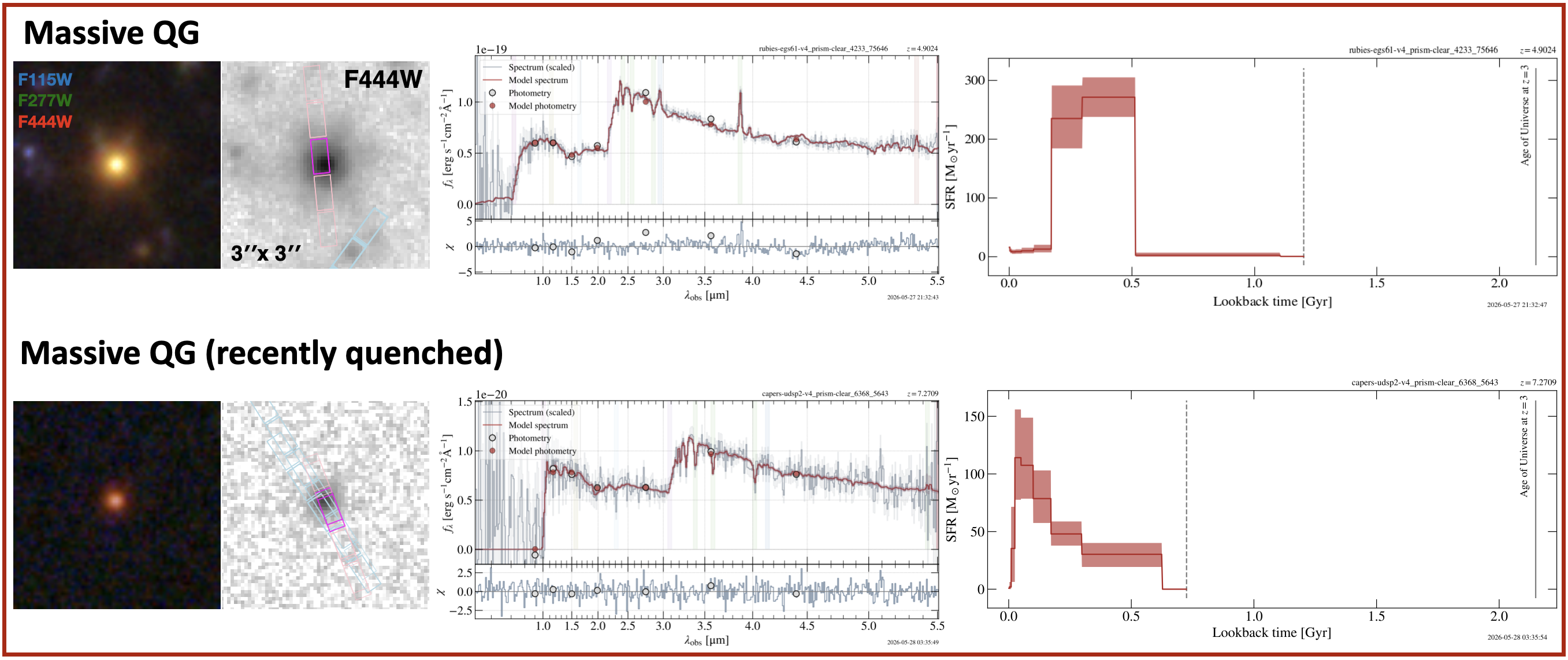} \hspace{-2mm}
\includegraphics[width=17.5cm]{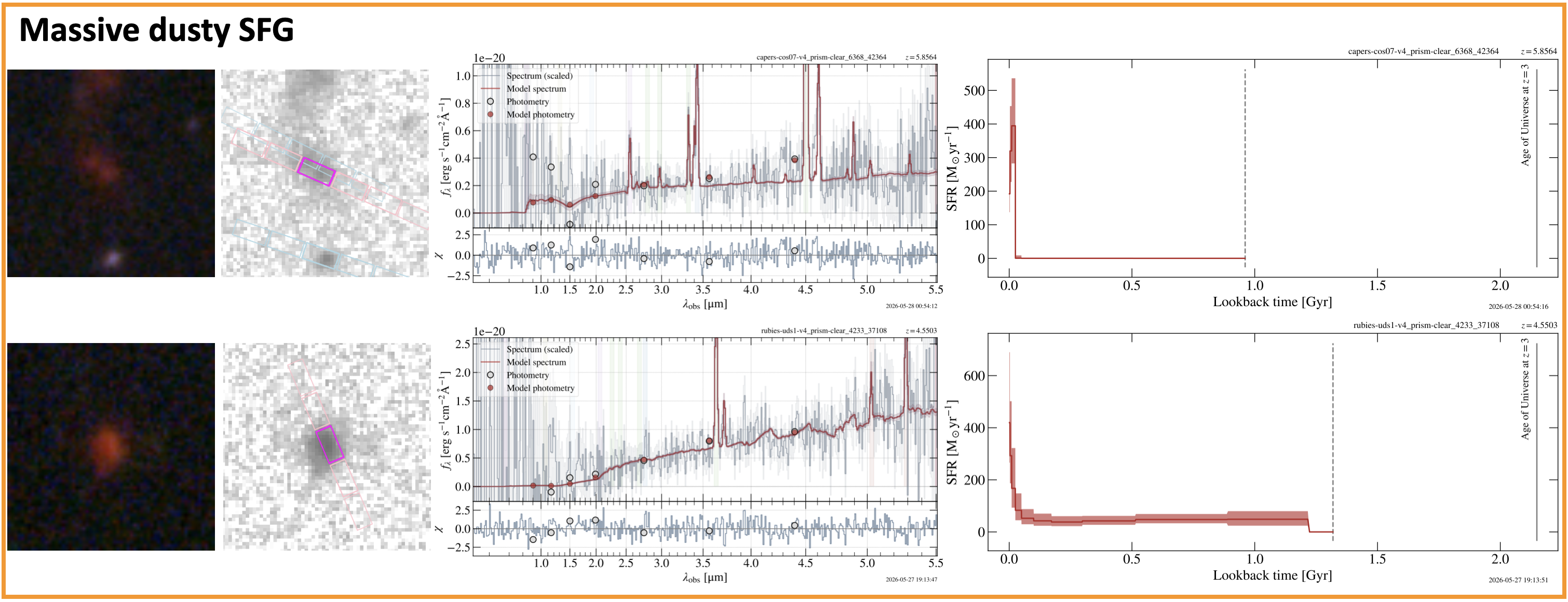}
\hspace{-2mm}
\includegraphics[width=17.5cm]{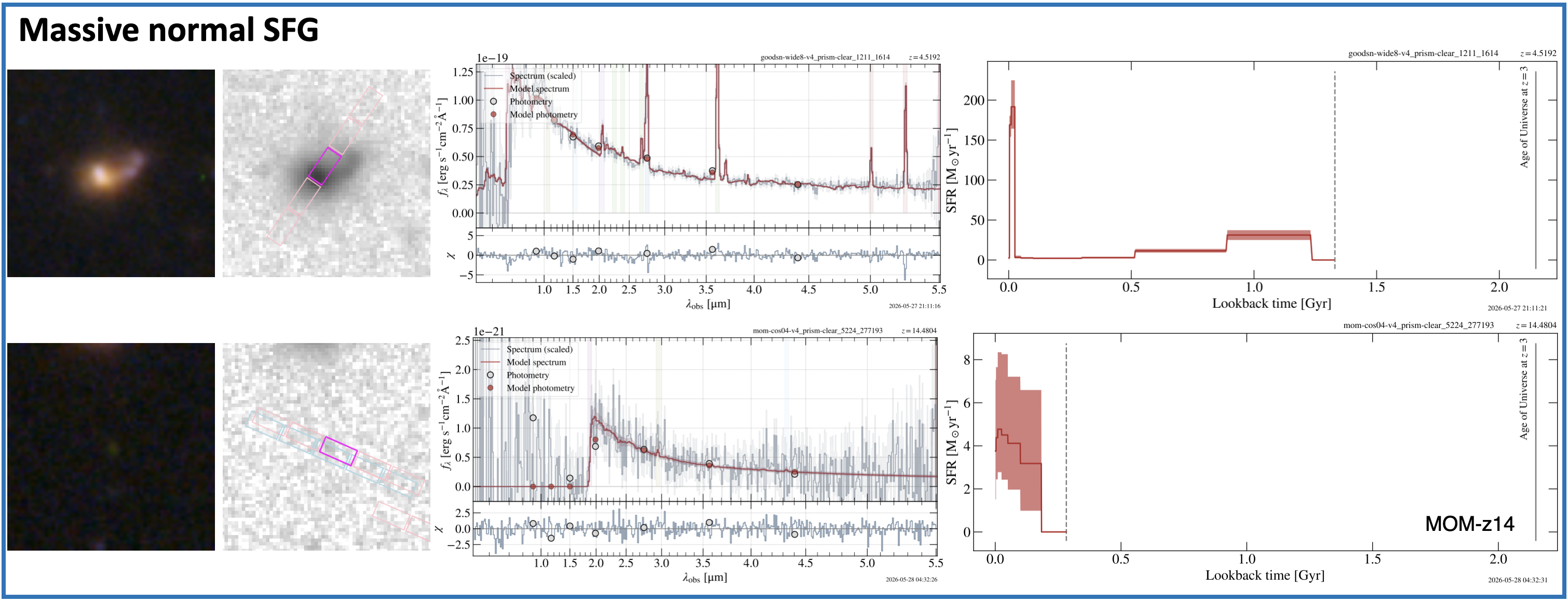}
\caption{\textbf{Examples of joint SED fitting of photometry and prism spectra (see Sect.~\ref{Sect:SED}) of our massive galaxies.} From left to right, we show the $3^{\prime\prime}\times 3^{\prime\prime}$ RGB image (F115W in blue, F277W in green, and F444W in red), the F444W image stamp, the best-fit \texttt{Bagpipes} model (red curve) together with the photometry (grey points) and prism spectrum (grey curve), and the inferred SFH. From top to bottom, we show: a massive QG at $z_{\rm spec}=4.90$ \citep{deGraaff2025}; a recently quenched massive QG at $z_{\rm spec}=7.27$ \citep{Weibel2024b}; two massive dusty SFGs, MAMBO-9-B at $z_{\rm spec}=5.86$ \citep{Akins2025_MAMBO9} and UDS-86 at $z_{\rm spec}=4.55$ (Qi et al. in prep.); and two massive normal SFGs, a source in the GOODS-North field at $z_{\rm spec}=4.52$ and MoM-z14 at $z_{\rm spec}=14.48$ \citep{Naidu2026_momz14}. All SED-fitting products for the full parent sample of prism spectra, from which the massive galaxies analyzed in this work are drawn, are publicly available in \cite{gottumukkala2026}.
}
\label{fig:SED}
\end{figure*}
\clearpage
\section{Robustness of stellar mass: with and without MIRI. }\label{appendix1} 
To evaluate the impact of mid-infrared photometry on the derived stellar masses, we compare \texttt{Bagpipes} fits performed with and without MIRI photometry for all galaxies with available MIRI observations. Since AGN emission is not included in our fiducial \texttt{Bagpipes} model, only MIRI filters up to F1000W are used in the fitting to minimize contamination from hot dust emission. To isolate the effect of MIRI photometry, all other fitting methods and assumptions, including joint SED fitting of photometry and prism spectra, the non-parametric SFH, BPASS stellar population models, the \cite{Salim2018} dust attenuation curve, are kept identical to the fiducial setup adopted throughout this work.

We require each source to have at least one MIRI band detected at S/N $>$ 3. This yields 1,021 of the 6,312 galaxies in the parent prism sample and 61 of the 200 galaxies in our massive galaxy sample with MIRI detections.
Details of the MIRI catalog used in our work are shown as follows:
\begin{itemize}
    \item SMILES \citep[][]{Alberts2024, Rieke2024}: covers an area of $\sim$34 arcmin$^2$ in the GOODS-S field with eight MIRI bands spanning 5.6-25.5$\mu$m (F560W, F770W, F1000W, F1280W, F1500W, F1800W, F2100W, and F2550W).
    \item COSMOS-Web MIRI \citep[][]{Shuntov2025catalog, Harish2025}: covers a non-contiguous area of $\sim$0.2 deg$^2$ in the COSMOS-Web field with the MIRI/F770W filter. The typical 5$\sigma$ point-source depth is $\sim$25.51 mag.
    \item MIDIS \citep[][]{Ostlin2025}: covers a deep area of $\sim$2.4 arcmin$^2$ in the HUDF with the MIRI/F560W filter. The typical 5$\sigma$ point-source depth is 28.65 mag.
    \item MINERVA UDS \citep[][]{Muzzin2025}: covers an area of $\sim$125 arcmin$^2$ in the UDS field with the MIRI F770W, F1280W, F1500W, and F1800W filters, where the F770W and F1800W data are provided by the PRIMER survey (PI: Dunlop). The typical 5$\sigma$ point-source depth is $\sim$23.9 mag in F1280W.
\end{itemize}

\begin{figure*}[h!]
\centering
\includegraphics[width=18.5cm]{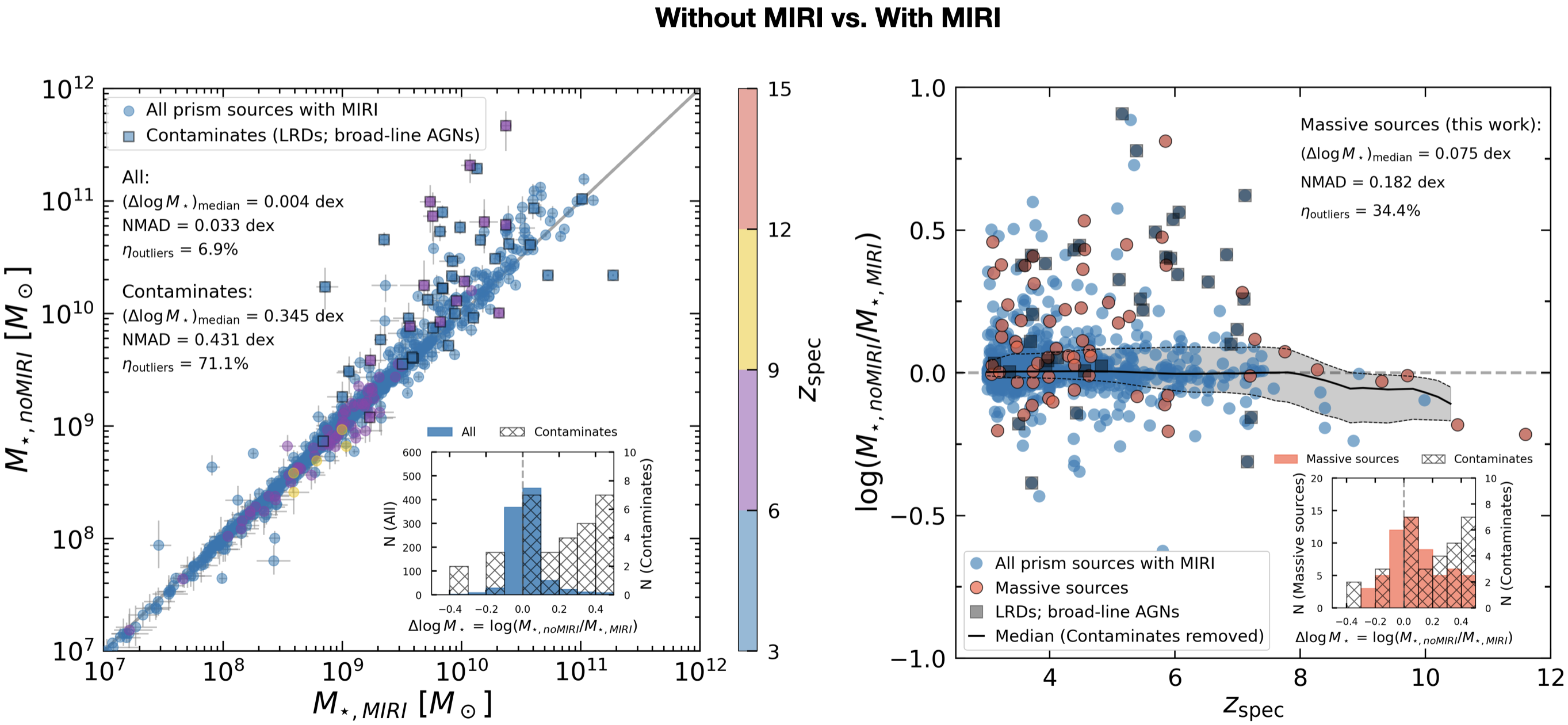}
\caption{\textbf{Comparison of stellar mass estimates with and without MIRI.} The squares indicate likely contaminants, including LRDs and broad-line AGNs, whose stellar mass estimates are considered unreliable and have therefore been excluded from our final massive galaxy catalog (see Sect.~\ref{Sect:agn}). In the right panel, the black solid line and the shaded region show the sliding median and the corresponding 16th-84th percentile range for all prism sources with MIRI photometry after excluding contaminants. The massive galaxies studied in this work are highlighted in red, with the derived median $\Delta\log M_\star$, NMAD, and outlier fraction indicated. Owing to the strong constraints provided by the joint SED fitting of photometry and prism spectroscopy, we find no significant systematic offset (median offset <0.1 dex) between the stellar masses derived with and without MIRI photometry for the massive galaxies in our sample.
}
         \label{compare_miri}
\end{figure*}

Figure~\ref{compare_miri} compares the stellar masses derived from \texttt{Bagpipes} with and without MIRI photometry.  We quantify the agreement using the median offset,
\begin{equation}
\Delta\log M_\star = \log(M_{\star,\rm noMIRI}/M_{\star,\rm MIRI}),
\end{equation}
the normalized median absolute deviation (NMAD),
\begin{equation}
{\rm NMAD}
=1.4826\times
{\rm median}\left(
\left|
\Delta\log M_\star-
{\rm median}(\Delta\log M_\star)
\right|
\right),
\end{equation}
and the outlier fraction, defined as the fraction of sources with
$|\Delta\log M_\star|>0.2$ dex.

\clearpage
\section{Robustness of stellar mass: under different modeling assumptions.}\label{appendix2} 
\begin{figure*}[h!]
\centering
\includegraphics[width=18cm]{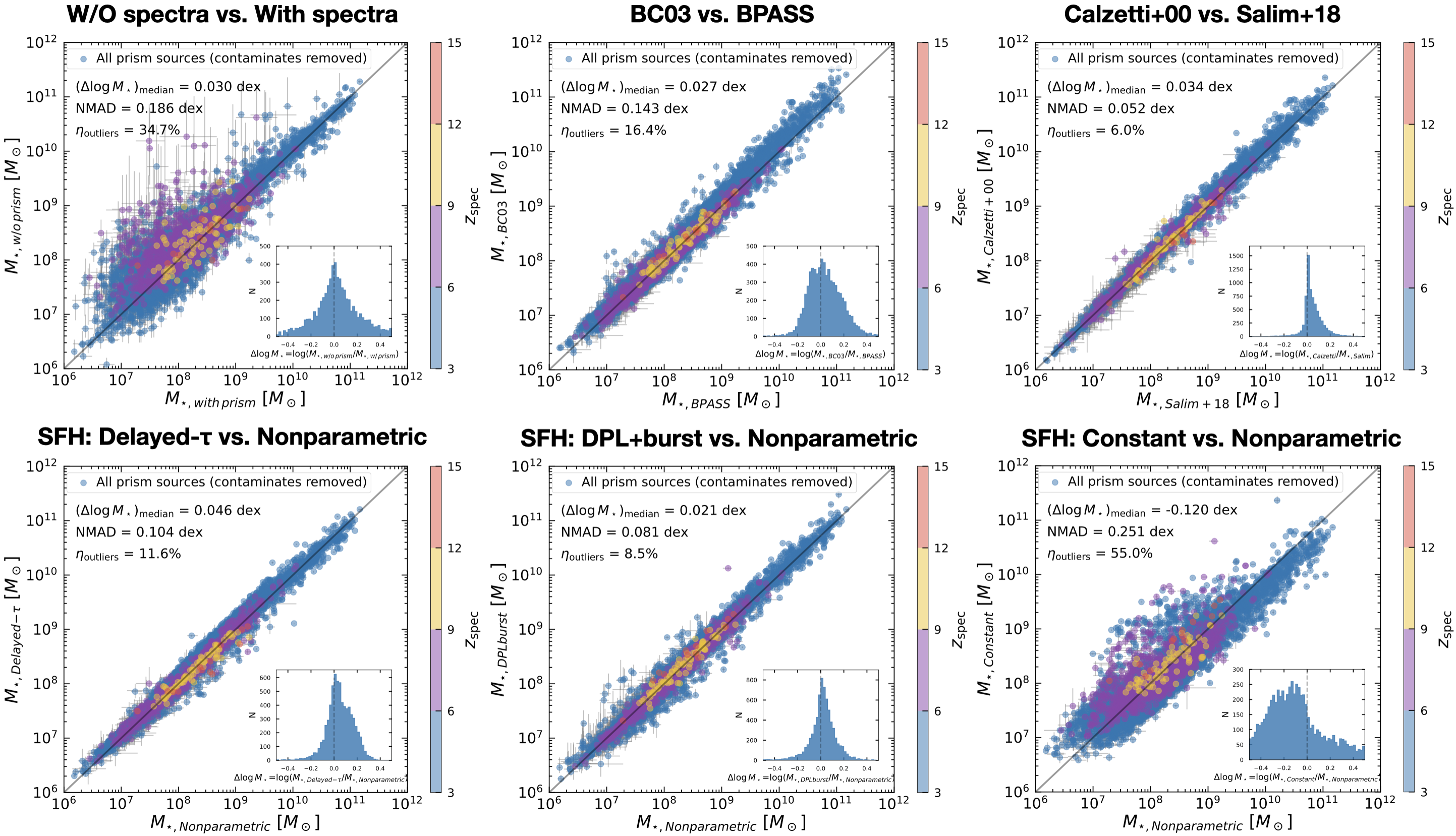}
\caption{\textbf{Comparison of stellar mass estimates under different \texttt{Bagpipes} modeling assumptions.} 
From left to right and top to bottom, the panels compare stellar masses derived using 1) photometry only vs. photometry+prism spectroscopy, 2) BC03 vs. BPASS stellar population models, 3) the  \cite{Calzetti2000} vs. \cite{Salim2018} dust attenuation curves, and three alternative SFH relative to our fiducial non-parametric SFH: 4) delayed-$\tau$ SFH; 5) double-power-law (DPL) + burst SFH; 6) constant SFH. Likely contaminants, including LRDs and broad-line AGNs (see Sect.~\ref{Sect:agn}), have been excluded from these comparisons.
}
         \label{compare}
\end{figure*}

To assess the dependence of our stellar mass estimates on the adopted SED modeling assumptions, we repeat the \texttt{Bagpipes} fitting using several alternative configurations. Specifically, we compare the effects of using (1) photometry only vs. photometry+prism spectroscopy, (2) BC03 vs. BPASS stellar population synthesis models, (3) the \cite{Calzetti2000} vs. \cite{Salim2018} dust attenuation curves, and (4) three alternative SFHs (delayed-$\tau$, double-power-law + burst, and constant SFH) relative to our fiducial non-parametric SFH. Likely contaminants, including LRDs and broad-line AGNs, are excluded from these comparisons. Except for the photometry-only case, all fits are performed using joint SED fitting of photometry and prism spectroscopy.
The median offsets, NMAD values, and outlier fractions for each comparison are computed following the same methodology described in Appendix~\ref{appendix1} and are reported in the corresponding panels of Fig.~\ref{compare}.
These statistics are summarized in Tables~\ref{tab:robustness} and \ref{tab:robustness_massive} for the full spectroscopic sample and the 200 most massive galaxies, respectively.

\begin{table*}[h!]
\caption{Comparison of stellar masses derived using different SED modeling assumptions relative to our fiducial \texttt{Bagpipes} fitting.}
\centering
\renewcommand{\arraystretch}{1.05}
\begin{tabular}{lcccc}
\hline\hline
Alternative fitting assumption &
Median offset &
NMAD &
Outlier fraction  &
Massive galaxies$^{a}$ \\
 &
(dex) &
(dex) &
(\%) & 
remaining (\%) \\
\hline
With MIRI       & 0.004   & 0.033   & 6.9  & 55/61 (90\%)$^{b}$ \\
Photometry only         &  0.030  & 0.186  & 34.7   &  165/200 (83\%) \\
BC03           & 0.027  & 0.143  & 16.4  & 173/200 (87\%)  \\
\cite{Calzetti2000} attenuation curve &  0.034 & 0.052  & 6.0  & 176/200 (88\%)   \\
Delayed-$\tau$ SFH      & 0.046  & 0.104  & 11.6  & 153/200 (77\%)   \\
double-power-law + burst SFH           & 0.021  & 0.081  & 8.5  &  165/200 (83\%)  \\
Constant SFH            & -0.120  & 0.251  & 55.0  & 98/200 (49\%)  \\
\hline
\end{tabular}

\tablefoot{
All stellar-mass differences are measured relative to our fiducial \texttt{Bagpipes} fitting (Sect.~\ref{Sect:SED}), which adopts joint photometry+prism fitting, the BPASS stellar population synthesis models, the \cite{Salim2018} dust attenuation curve, and a non-parametric SFH. The median offset, NMAD, and outlier fraction are computed for the full spectroscopic sample used in each comparison.\\
$^{a}$ Fraction of the 200 most massive galaxies that remain above the adopted evolving stellar-mass threshold (Sect.~\ref{Sect:massive sample}) under each alternative fitting assumption. $^{b}$ Among the 200 most massive galaxies, 61 have available MIRI photometry, and 55 sources remain above the adopted evolving stellar-mass threshold after adding MIRI photometry.
}
\label{tab:robustness}
\end{table*}

\begin{table*}[h!]
\caption{Comparison of stellar masses for the 200 most massive galaxies derived using different SED modeling assumptions relative to our fiducial \texttt{Bagpipes} fitting.}
\centering
\renewcommand{\arraystretch}{1.05}
\begin{tabular}{lcccc}
\hline\hline
Alternative fitting assumption &
Median offset &
NMAD &
Outlier fraction  \\
 &
(dex) &
(dex) &
(\%) \\
\hline
With MIRI       & 0.075   & 0.182   & 34.4   \\
Photometry only  & -0.142 & 0.179 & 40.1   \\
BC03           &  -0.052 & 0.151 & 29.4 \\
\cite{Calzetti2000} attenuation curve & -0.067 & 0.110 & 14.1    \\
Delayed-$\tau$ SFH   & -0.152 & 0.131 & 36.4       \\
double-power-law + burst SFH   & -0.102 & 0.117 & 22.7          \\
Constant SFH    & -0.351 & 0.225 & 74.2          \\
\hline
\end{tabular}

\tablefoot{
The fitting assumptions and statistical definitions follow Table~\ref{tab:robustness}, but for the 200 most massive galaxies. For the comparison with MIRI, only the 61 massive galaxies with available MIRI photometry are included.}
\label{tab:robustness_massive}
\end{table*}

\section{Robustness of SFR: under different modeling assumptions.}\label{appendix3}

\begin{figure*}[h!]
\centering
\includegraphics[width=17.5cm]{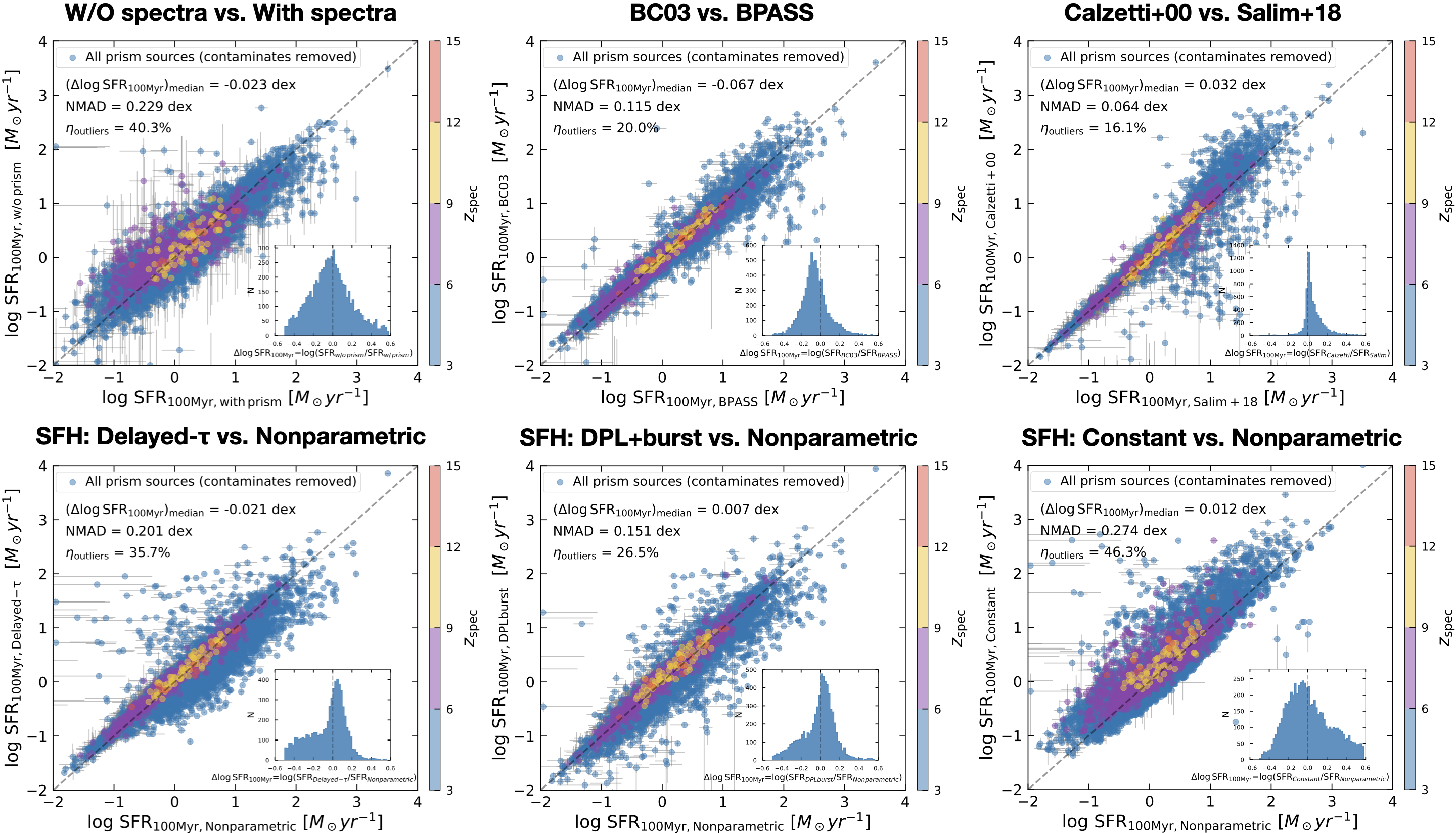}
\caption{\textbf{Comparison of SFR$_{\rm 100\,Myr}$ estimates under different \texttt{Bagpipes} modeling assumptions.} 
Similar to Fig.~\ref{compare}, but for SFR$_{\rm 100\,Myr}$.
}
         \label{compare_SFR100}
\end{figure*}

\begin{figure*}[h!]
\centering
\includegraphics[width=17.5cm]{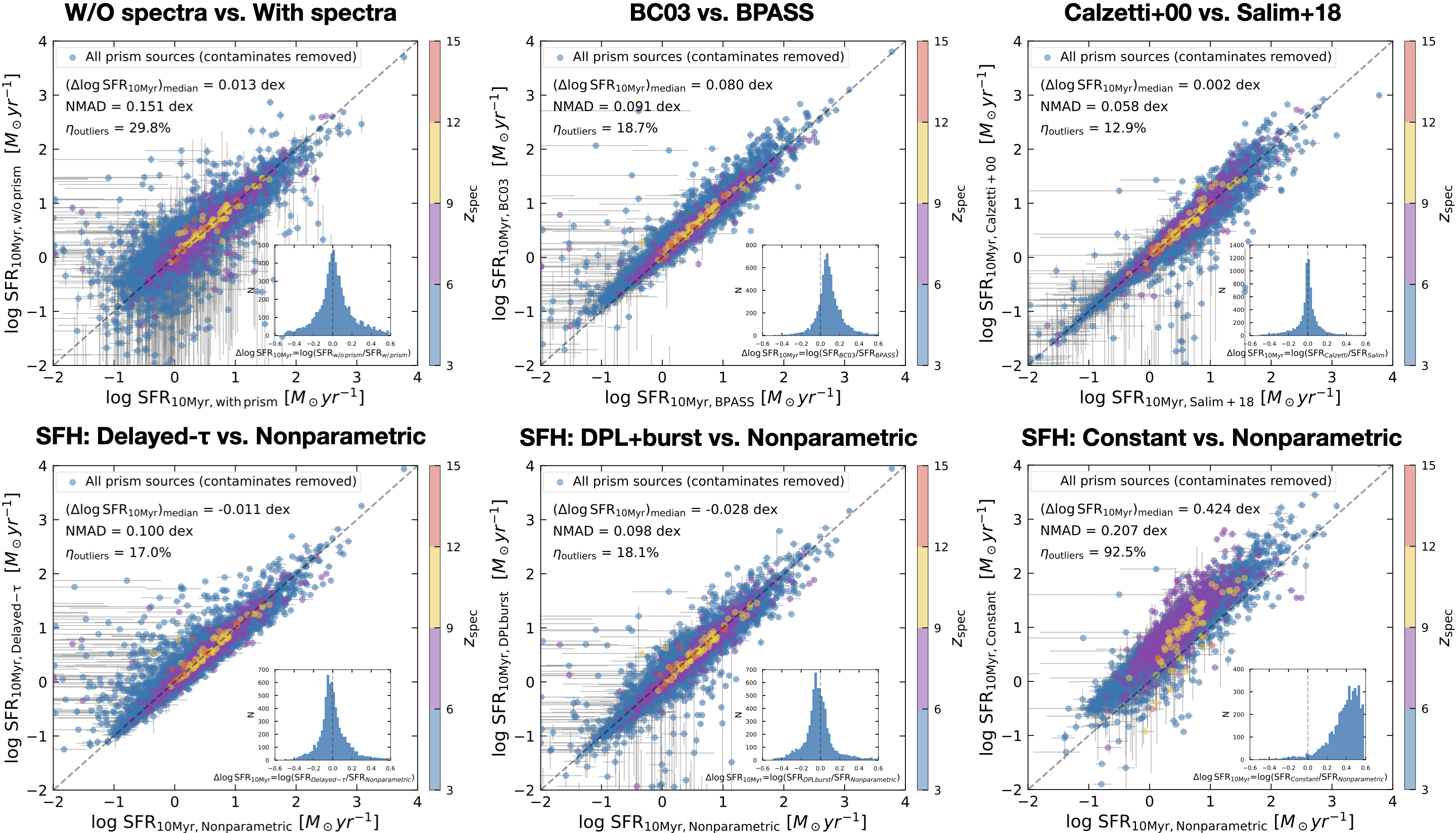}
\caption{\textbf{Comparison of SFR$_{\rm 10\,Myr}$ estimates under different \texttt{Bagpipes} modeling assumptions.} 
Similar to Fig.~\ref{compare}, but for SFR$_{\rm 10\,Myr}$.
}
         \label{compare_SFR10}
\end{figure*}

Similar to Appendix~\ref{appendix2}, here we assess the dependence of SFR$_{\rm 100\,Myr}$ and SFR$_{\rm 10\,Myr}$ estimates on the adopted SED modeling assumptions.
The median offsets, NMAD values, and outlier fractions for the full spectroscopic sample for each comparison  are reported in the corresponding panels of Figs.~\ref{compare_SFR100} and \ref{compare_SFR10} and summarized in Table~\ref{tab:robustness_sfr}.
The corresponding statistics for the 200 most massive galaxies are listed in Table~\ref{tab:robustness_sfr_massive}.

For the full spectroscopic sample, excluding the intentionally simplified constant-SFH model, the median offsets are small ($<0.1$ dex in absolute value) for both SFR$_{\rm 100\,Myr}$ and SFR$_{\rm 10\,Myr}$.
The NMAD ranges from 0.064 to 0.229 dex and from 0.058 to 0.151 dex, respectively.
For the massive sample, SFR$_{\rm 100\,Myr}$ shows larger median offsets, ranging from $-0.227$ to $-0.139$ dex, with NMAD values of 0.233-0.406 dex.
In comparison, the median offsets in SFR$_{\rm 10\,Myr}$ remain small ($<0.1$ dex in absolute value), although the NMAD values range from 0.144 to 0.264 dex.

The constant-SFH model produces the largest systematic offset in SFR$_{\rm 10\,Myr}$, with median offsets of 0.424 and 0.467 dex for the full and massive samples, respectively.
As discussed in Sect.~\ref{Sect:robust}, this model is included only as a simplified limiting case, since it cannot reproduce bursty star formation or rapid quenching. For this model, SFR$_{\rm 10\,Myr}$ and SFR$_{\rm 100\,Myr}$ are identical when constant star formation extends over at least the last 100 Myr.
Overall, the massive sample shows a stronger modeling dependence than the full sample, particularly for SFR$_{\rm 100\,Myr}$.
The corresponding effects on galaxy classification are discussed in Appendix~\ref{appendix_classification}.


\begin{table*}[h!]
\caption{Comparison of SFR$_{\rm 100\,Myr}$ and SFR$_{\rm 10\,Myr}$ derived using different SED modeling assumptions relative to our fiducial \texttt{Bagpipes} fitting.}
\centering
\renewcommand{\arraystretch}{1.05}
\begin{tabular}{lcccccc}
\hline\hline
 & \multicolumn{3}{c}{SFR$_{\rm 100\,Myr}$}
 & \multicolumn{3}{c}{SFR$_{\rm 10\,Myr}$} \\
\cmidrule(lr){2-4}\cmidrule(lr){5-7}
Alternative fitting assumption &
Median offset &
NMAD &
Outlier fraction & Median offset &
NMAD &
Outlier fraction  \\
 &
(dex) &
(dex) &
(\%) & 
(dex) &
(dex) &
(\%)
 \\
\hline
Photometry only         &  -0.023  & 0.229  & 40.3 & 0.013 & 0.151 & 29.8   \\
BC03                    & -0.067  & 0.115  & 20.0  & 0.080 & 0.091 & 18.7  \\
\cite{Calzetti2000} attenuation curve &  0.032 & 0.064  & 16.1 & 0.002 & 0.058 & 12.9  \\
Delayed-$\tau$ SFH      & -0.021  & 0.201  & 35.7 & -0.011 & 0.100 & 17.0   \\
double-power-law + burst SFH           & 0.007  & 0.151  & 26.5 & -0.028 & 0.098 & 18.1   \\
Constant SFH            & 0.012  & 0.274  & 46.3 & 0.424 & 0.207 & 92.5   \\
\hline
\end{tabular}

\tablefoot{
All SFR differences are measured relative to our fiducial \texttt{Bagpipes} fitting (Sect.~\ref{Sect:SED}), which adopts joint photometry+prism fitting, the BPASS stellar population synthesis models, the \cite{Salim2018} dust attenuation curve, and a non-parametric SFH. The median offset, NMAD, and outlier fraction are computed for the full spectroscopic sample used in each comparison.}
\label{tab:robustness_sfr}
\end{table*}

\begin{table*}[h!]
\caption{Comparison of SFR$_{\rm 100\,Myr}$ and SFR$_{\rm 10\,Myr}$ for the 200 most massive galaxies derived using different SED modeling assumptions relative to our fiducial \texttt{Bagpipes} fitting.}
\centering
\renewcommand{\arraystretch}{1.05}
\begin{tabular}{lcccccc}
\hline\hline
 & \multicolumn{3}{c}{SFR$_{\rm 100\,Myr}$}
 & \multicolumn{3}{c}{SFR$_{\rm 10\,Myr}$} \\
\cmidrule(lr){2-4}\cmidrule(lr){5-7}
Alternative fitting assumption &
Median offset &
NMAD &
Outlier fraction & Median offset &
NMAD &
Outlier fraction  \\
 &
(dex) &
(dex) &
(\%) & 
(dex) &
(dex) &
(\%)
 \\
\hline
Photometry only         & -0.227 & 0.328 & 68.0 & 0.004 & 0.237 & 39.1   \\
BC03                    & -0.139 & 0.233 & 56.3 & 0.081 & 0.144 & 34.5 \\
\cite{Calzetti2000} attenuation curve & -0.156 & 0.406 & 57.1 & -0.004 & 0.264 & 46.5 \\
Delayed-$\tau$ SFH      &  -0.180 & 0.326 & 64.1 & 0.005 & 0.151 & 37.4  \\
double-power-law + burst SFH & -0.193 & 0.380 & 59.6 & -0.043 & 0.145 & 30.8 \\
Constant SFH            & 0.105 & 0.445 & 62.9 & 0.467 & 0.291 & 85.1 \\
\hline
\end{tabular}

\tablefoot{
The fitting assumptions and statistical definitions follow Table~\ref{tab:robustness_sfr}, but for the 200 most massive galaxies.}
\label{tab:robustness_sfr_massive}
\end{table*}

\clearpage
\section{Classification of the 200 massive galaxies under different modeling assumptions.}\label{appendix_classification}

We assess the dependence of the classification of the 200 massive galaxies on the adopted SED modeling assumptions.
The populations of normal SFGs, dusty SFGs, and QGs follow the definitions in Sect.~\ref{Sect:clean massive}, with QGs including recently quenched galaxies.
Table~\ref{tab:classification} lists the number of galaxies in each population under each fitting configuration, together with the overlap with the corresponding fiducial population.
The same 200 galaxies are used in all comparisons, irrespective of whether their revised stellar masses remain above the adopted mass threshold.

The classifications of normal and dusty SFGs are comparatively stable, with 81-101 (77-96\%) of the 105 fiducial normal SFGs and 53-66 (80-100\%) of the 66 fiducial dusty SFGs remaining in their original populations.
The QG classification shows a stronger dependence on the adopted assumptions, particularly the SFH.
The photometry-only, BC03, and Calzetti configurations recover 18 (62\%), 22 (76\%), and 29 (100\%) of the 29 fiducial QGs, respectively.
In comparison, the delayed-$\tau$, double-power-law + burst, and constant-SFH models recover 6 (21\%), 20 (69\%), and 10 (34\%) fiducial QGs, respectively. Besides the known limitations of the constant-SFH model, the particularly low recovery of QGs with the delayed-$\tau$ model may reflect its limited flexibility: the same parameter controls both the time to peak star formation and the subsequent decline, making it difficult to describe extended star formation followed by rapid quenching. 
Thus, although most normal and dusty SFGs retain their original classification, the inferred number and membership of the QG population are more sensitive to the adopted SFH parameterization and priors.

\begin{table*}[h!]
\caption{Classification of the 200 most massive galaxies under different SED modeling assumptions.}
\centering
\renewcommand{\arraystretch}{1.05}
\begin{tabular}{lccc}
\hline\hline
Alternative fitting assumption &
Normal SFGs &
Dusty SFGs &
QGs  \\
 & $N$ ($N_{\rm remaining}$) & $N$ ($N_{\rm remaining}$) & $N$ ($N_{\rm remaining}$) \\
\hline
Fiducial$^{*}$      & 105 (105) & 66 (66) & 29 (29) \\
Photometry only         & 123 (101) & 58 (55) & 19 (18)  \\
BC03           & 102 (94) & 73 (62) & 25 (22)  \\
\cite{Calzetti2000} attenuation curve & 83 (81)  & 82 (62) & 35 (29)   \\
Delayed-$\tau$ SFH    & 130 (96) & 64 (54) & 6 (6)     \\
double-power-law + burst SFH   & 115 (94) & 64 (53) & 21 (20)          \\
Constant SFH      & 101 (84) & 88 (66) &  11 (10)       \\
\hline
\end{tabular}

\tablefoot{
Numbers give the total number of galaxies in each population under each fitting assumption. Numbers in parentheses give the overlap with the same population in the fiducial classification. The same 200 galaxies are used in all comparisons. $^{*}$ Fiducial \texttt{Bagpipes} fitting used in this paper (Sect.~\ref{Sect:SED}), which adopts joint photometry+prism fitting, the BPASS stellar population synthesis models, the \cite{Salim2018} dust attenuation curve, and a non-parametric SFH.}
\label{tab:classification}
\end{table*}

\section{Catalog of 200 massive galaxies.}\label{appendix_table1} 

\begin{table*}[h!]
\caption{Programs of 200 massive galaxies among 264 JWST/NIRSpec prism observations.}   
\tiny          
\centering
\renewcommand{\arraystretch}{1.2} 
\begin{threeparttable} 
 
\begin{tabular}{l c  }    
\hline\hline       
                    
Programs  & spectra\\ 
\hline  
CAPERS (GO-6368; PI: Dickinson)&   52    \\
CEERS-ERS (GO-1345; \citealt{Finkelstein2025})&   5    \\
DD-2750 (\citealt{Arrabal_Haro2023})&   5    \\
DD-6585 (PI: Coulter)&   1    \\
GO-2198 (\citealt{Barrufet2025})&   6    \\
GO-2565 (\citealt{Nanayakkara2025}) & 17\\
GO-4106 (PI: Nelson)&   1    \\
GO-5997 (PI: Looser)&   2    \\
GO-8018 (PI: Lin)&   2    \\
GO-8060 (PI: Egami)&   3    \\
GTO-WIDE (GTO-1211; GTO-1212; GTO-1213; GTO-1214; GTO-1215; \citealt{Maseda2024}) & 35\\
JADES (GTO-1180; GTO-1181; GTO-1210; GTO-1286; GTO-1287; GO-3215; \citealt{D_Eugenio2024}; \citealt{Eisenstein2023}) & 45\\
MoM (GO-5224; PIs: Naidu \& Oesch; \citealt{Naidu2026_momz14}) & 18\\
RUBIES (GO-4233; \citealt{deGraaff2025_rubies}) &   74    \\
\hline
\end{tabular}
\label{surveys} 
\end{threeparttable} 
\end{table*}

\begin{table*}[h!]
\caption{Physical properties of the 200 most massive galaxies at $z>3$.}   
\centering
\renewcommand{\arraystretch}{1.5} 
\begin{threeparttable} 
 
\begin{tabular}{c c c c c c c c c}    
\hline\hline       
                    
RA & Dec & $z_{\rm spec}$  & log$(M_{\star}/M_\odot)$ & log(SFR$_{100}$/$M_{\odot}$yr$^{-1}$) & log(SFR$_{10}$/$M_{\odot}$yr$^{-1}$) & $A_{\rm V}$ [mag] & Classification & ... \\ 
\hline   
189.313126 & 62.159092 & 3.0124 & $10.99_{-0.01}^{+0.01}$ & $-0.10_{-0.20}^{+0.17}$ & $0.38_{-0.07}^{+0.07}$ & $0.58_{-0.03}^{+0.02}$ & QG & ...\\
\hline
\end{tabular}
\begin{tablenotes}
\item \textbf{Note.} The whole catalog and the description of its contents can be found at \url{https://doi.org/10.5281/zenodo.23888775}.
\end{tablenotes}
\label{individual} 
\end{threeparttable} 
\end{table*}

\begin{figure*}[h!]
\centering
\includegraphics[width=10cm]{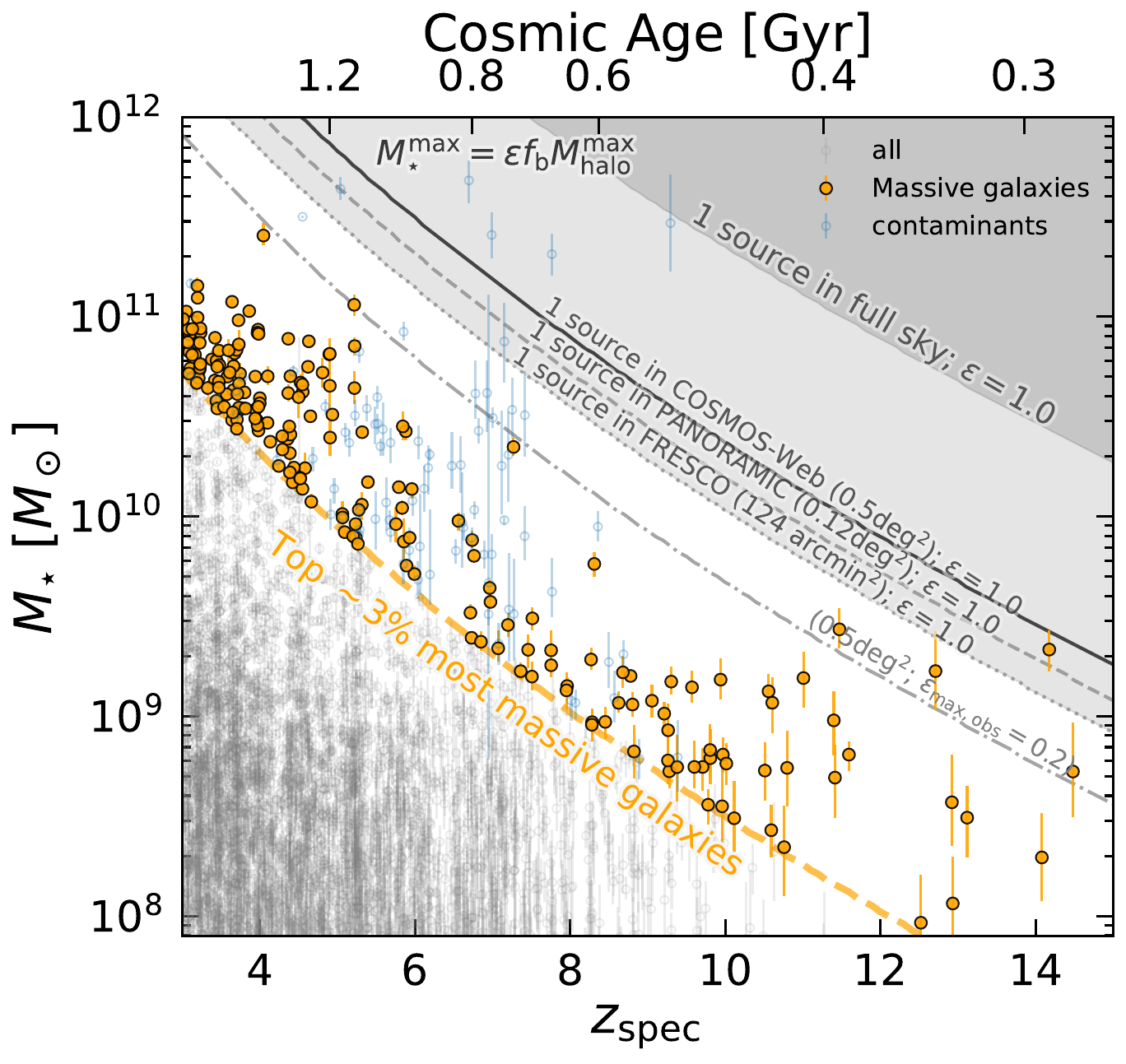}
\caption{\textbf{Similar to Fig.~\ref{fig2}, but shows the locations of contaminants, where their stellar mass are largely uncertain}. 
Orange points represent the top $\sim$3\% (200) most massive galaxies among all 6,312 galaxies (grey open points) at $z_{\rm spec}>3$ with JWST/NIRSpec prism observations in the COSMOS, UDS, EGS, and GOODS fields. Blue open points indicate likely contaminants from broad-line AGNs or LRDs, whose stellar masses are considered unreliable and have therefore been removed from our final massive galaxy catalog (see Sect.~\ref{Sect:agn}). 
}
         \label{fig2_}
\end{figure*}

\begin{figure*}
\centering
\includegraphics[width=8cm]{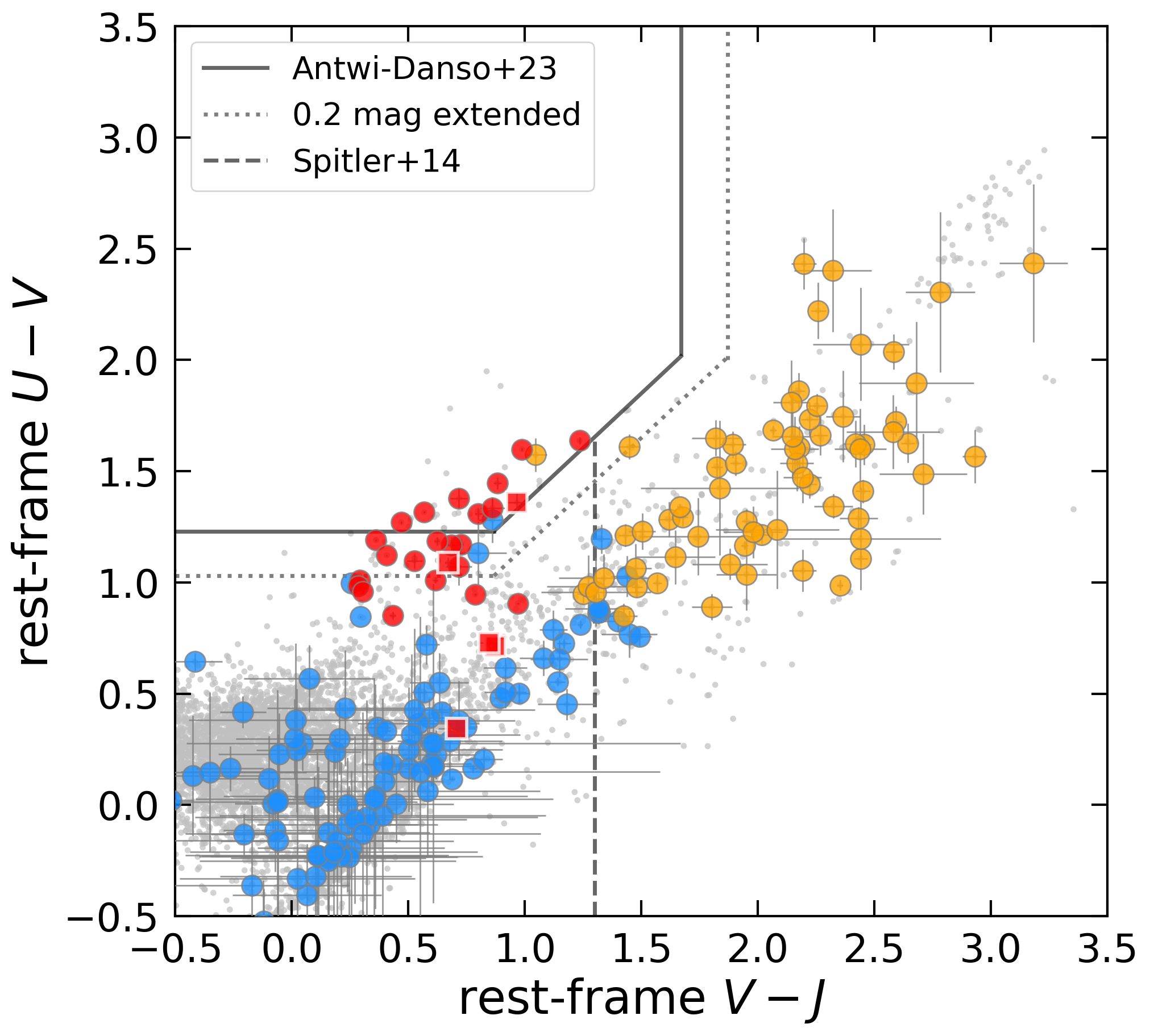} 
\includegraphics[width=8cm]{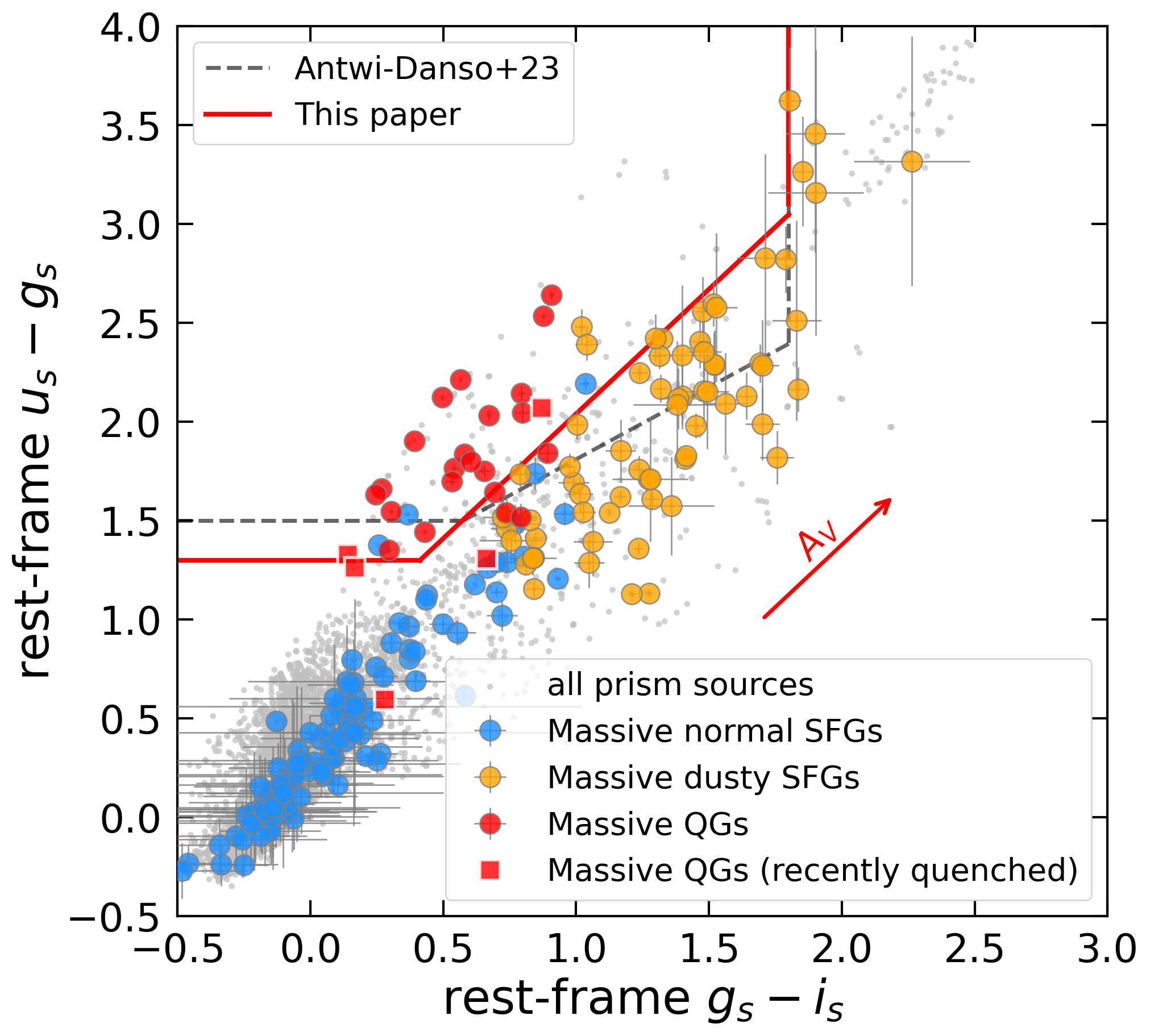}
\caption{\textbf{Comparison between rest-frame \textit{UVJ} (\textit{left}) and (\textit{ugi})$_{\rm s}$ (\textit{right}) selections and our spectroscopically classified massive galaxies. }
Similar to Fig.~\ref{uvj_ugis}, but using rest-frame colors derived from photometry-only SED fitting. The fits use only seven NIRCam broad-band bands (F090W, F115W, F150W, F200W, F277W, F356W, and F444W), while retaining the spectroscopic-redshift constraints.
The galaxy classifications and selection boundaries are unchanged from Fig.~\ref{uvj_ugis}.
}
\label{uvj_ugis_phot}
\end{figure*}

\end{appendix}

\end{document}